\documentclass[bibyear]{aa}  
\usepackage[justification=centering]{caption}
\usepackage[rightcaption]{sidecap}
\usepackage{subcaption}
\usepackage{graphicx}
\usepackage{natbib}
\usepackage{color}
\usepackage{threeparttable}
\usepackage[singlelinecheck=false,justification=justified]{caption}
\usepackage{txfonts}

\usepackage{gensymb}

\begin{document}

    
    \title{Stellar obliquity measurements of six gas giants}
    \subtitle{Orbital misalignment of WASP-101b and WASP-131b}
    \titlerunning{Stellar obliquity measurements of six gas giants}
    
   \author{J. Zak
          \inst{1, 2, 3}\fnmsep
          \and
          A. Bocchieri\inst{4}
          \and
          E. Sedaghati\inst{5}
           \and
          H.\,M.\,J. Boffin\inst{1}
           \and
          Z. Prudil\inst{1}
           \and
          M. Skarka\inst{2}
            \and
          Q. Changeat\inst{6}
            \and\\
          E. Pascale\inst{4}
           \and
          D. Itrich \inst{1, 7}
          \and
          V. D. Ivanov \inst{1}
          \and
          M. Vitkova \inst{2, 8}
          \and
          P. Kabath
          \inst{2}
          \and
          M. Roth\inst{3, 9}
                \and
          A. Hatzes\inst{3, 9}
          }

   \institute{European Southern Observatory, Karl-Schwarzschild-str. 2, 85748 Garching, Germany\\
              \email{jiri.zak@eso.org}
         \and
         Astronomical Institute of the Czech Academy of Sciences, Fri\v{c}ova 298, 25165 Ond\v{r}ejov, Czech Republic
         \and
        Faculty of Physics and Astronomy, Friedrich-Schiller-Universität, Fürstengraben 1, 07743, Jena, Germany
        \and
             Dipartimento di Fisica, La Sapienza Università di Roma, Piazzale Aldo Moro 5, Roma, 00185, Italy
        \and
             European Southern Observatory, Casilla 13, Vitacura, Santiago, Chile 
        \and
             European Space Agency (ESA), ESA Office, Space Telescope Science Institute (STScI), Baltimore, MD 21218, USA.
        \and
Steward Observatory, The University of Arizona, Tucson, AZ 85721, USA
        \and
Department of Theoretical Physics and Astrophysics, Masaryk University, Kotlá\v{r}ská 2, CZ-61137 Brno, Czech Republic
         \and
         Thueringer Landessternwarte Tautenburg, Sternwarte 5, 07778 Tautenburg, Germany
             }

   \date{Received September 15, 1996; accepted March 16, 1997}

 
  \abstract
   {One can infer the orbital alignment of exoplanets with respect to the spin of their host stars using the Rossiter-McLaughlin effect, thereby giving us the chance to test planet formation and migration theories and improve our understanding of the currently observed population. 
   We analyze archival HARPS and HARPS-N spectroscopic transit time series of six gas giant exoplanets on short orbits,
namely WASP-77 Ab, WASP-101b, WASP-103b, WASP-105b, WASP-120b and WASP-131b. 
   We find a moderately misaligned orbit for WASP-101b ($\lambda =34\degree\ \pm$ 3) and a highly misaligned orbit for WASP-131b ($\lambda =161\degree\ \pm$ 5), while the four remaining ones appear aligned: WASP-77 Ab ($\lambda =-8\degree\ ^{+19}_{-18}$), WASP-103b ($\lambda =2\degree\ ^{+35}_{-36}$), WASP-105b ($\lambda =-14\degree\ ^{+28}_{-24}$), and WASP-120b ($\lambda =-2\degree\ \pm$ 4).
   For WASP-77 Ab, we were able to infer its true orbital obliquity ($\Psi =48\degree\ ^{+22}_{-21}$).
   We additionally perform transmission spectroscopy of the targets in search of strong atomic absorbers in the exoatmospheres, but are unable to detect any features, most likely due to the presence of high-altitude clouds or Rayleigh scattering muting the strength of the features. Finally, we comment on future perspectives for studying these targets with the upcoming space missions to investigate the evolution and migration histories of these planets.}


   \maketitle
%

\section{Introduction} \label{sec:intro}

The formation and evolution of exoplanets and how this is linked to the properties of their host stars and the environments in which they formed are long-standing questions. One possible clue may come from measuring the relative angle of the rotation axis of the star with that of the planetary orbital angular momentum. The planets of the Solar System, for example, are almost co-planar, as these angles -- called stellar obliquities -- are all very small, leading to the idea of planet formation within a disc. The sky-projected stellar obliquity can be inferred through the observations of the Rossiter-McLaughlin (R-M) effect, which is a spectroscopic anomaly that occurs when the disc of a companion passes in front of the spinning star. 
As usual in astronomy, this effect is misnamed, as it was first described by \cite{1893AstAp..12..646H}, and then actually detected in the binary system $\beta$ Lyrae by \citet{ross24}. The same year, and using the methodology of Rossiter, \citet{mcl} applied it to a study of Algol, while the first application of the R-M effect to exoplanets was carried out by \citet{quel00} on HD 209458. For a more detailed review of this history, the reader is encouraged to refer to \citet{alb11,alb22}. At first, only Jupiter-mass planets were characterized, but later development of instrumentation allowed for the characterization of Neptunian \citep{bour18}, sub-Neptunian \citep{dalal19} and even terrestrial planets \citep{hir20,zhao22}. Such studies have revealed a complex architecture of exoplanetary systems with planetary orbits varying from well-aligned, to slightly misaligned, to planets on polar and even retrograde orbits.

\begin{table*}[htbp]
\captionsetup{justification=centering}
\caption{Observing logs for all six planets. The number in parenthesis represents the number of frames taken in-transit.}
\vspace{-.4cm}
\label{obs_logs}
\centering
\begin{center}
\begin{tabular}{lcccccc}
\hline
Target & Date & No.     & Exp.     & Airmass & Median & Instrument \\ 
     &~Obs~&~frames~& Time (s) & range    &  S/N    \\ 
\hline
WASP-77 A & 2018-11-05~&~21 (9)~& 900     & 1.48-1.08-1.33~&~31.2-55.4 & HARPS \\ 

WASP-101 & 2015-12-20~&~54 (18)~& 500     & 1.99-1.01-1.47~&~27.2-48.3& HARPS \\ 

WASP-103 & 2016-06-01~&~33 (11)~& 900     &2.02-1.08-2.30~&~15.3-21.9& HARPS-N\\  

& 2016-06-14~&~29 (10)~& 900 & 1.48-1.07-2.24~&~6.5-21.9& HARPS-N \\

WASP-105 & 2014-09-30~&~24 (14)~& 900     & 1.28-1.07-1.36~&~11.9-17.9& HARPS\\  

& 2014-12-10~&~20 (13)~& 600-1200 & 1.09-1.07-1.82~&~8.8-18.7& HARPS \\
 
WASP-120 & 2018-10-20~&~ 37 (21)~& 600 & 1.73-1.04-1.14~&~17.2-31.2& HARPS \\  
 
WASP-131 & 2015-06-06~&~ 37 (21)~& 600 & 1.13-1.00-1.87~&~36.2-53.7& HARPS \\  
\hline 
\end{tabular}

\end{center}
\end{table*}

The R-M effect is thus a valuable tool as it can provide insight into the planetary formation and evolution \citep{tri17}. It can be used to test theories of planetary migration, for example to understand if the planet formed in situ or if it has migrated to its current location.
Polar and retrograde orbits are indicative of violent history within the system, as such orbits seem to oppose the standard formation processes for the planetary systems where both the star and the planets form from the same rotating disc.
Several mechanisms have been proposed to explain these extreme orbits, such as long-term interactions with a binary companion \citep{bat12}, close fly-bys of other stars \citep{bat10}, the shift of the stellar rotation axis relative to the disc due to the effects of the stellar magnetic field \citep{lai11}, gravitational scattering among the planets \citep{chat08}, or long term perturbations due to companion caused by Kozai-Lidov mechanism \citep{fab07}. \citet{bres19} showed that retrograde orbits can also be the outcome of prograde flybys.

Despite the large number (over 150) of R-M effect measurements, there are no definitive trends in the planetary population when plotting the projected obliquity versus other parameters. However, suggestion of some trends were proposed;  \citet{alb12} and \citet{daw14} identified two populations of hot Jupiters, based on the effective temperature of the host star, with the division being at the Kraft break -- an abrupt decrease in the stars' average rotation rates at about 6200\,K \citep[][]{kraft67}. \citet{ric22} found evidence that warm Jupiters in single-star systems form more quiescently compared to their hot analogues. \citet{ham22} found that hot Jupiter host stars with larger-than-median obliquities are older than hot Jupiter host stars with smaller-than-median obliquities.

In this study, we present the projected obliquity measurements of six gas giants on short orbital periods around F-G-K dwarfs with previously unpublished HARPS and HARPS-N archival data.

\section{Data sets and their analyses} \label{sec:data}
We have used datasets for five exoplanets from ESO's HARPS instrument mounted at the 3.6-m telescope at La Silla, Chile. Furthermore, the data for one planet (WASP-103) were obtained with the HARPS-N instrument mounted at the 3.5-m TNG telescope at La Palma, Canary Islands. The HARPS data were downloaded from the ESO science archive\footnote{\url{http://archive.eso.org}} (program IDs 0102.C-0618, 096.C-0331, 094.C-0090, 0102.C-0319 and 095.C-0105), while the HARPS-N data (program ID A33TAC17) were obtained from the TNG archive\footnote{\url{http://archives.ia2.inaf.it/tng/}}. Table \ref{obs_logs} displays the properties of the data sets that were used. The downloaded data are fully reduced products as processed with the HARPS Data Reduction Software (DRS version 3.8). Each spectrum is provided as a merged 1D spectrum resampled onto a 0.001 nm uniform wavelength grid. The wavelength coverage of the spectra spans from 380 to 690 nm, with a resolving power of R\,$\approx$\,115\,000 corresponding to 2.7~km\,s$^{-1}$ per resolution element. The spectra are already corrected to the Solar system barycentric frame of reference. We further summarize the properties of the studied systems in Table \ref{tab:targets1}.

\begin{table*}[htbp]
\caption{Properties of the targets (star and planet).}
\vspace{-.4cm}
\label{tab:targets1}
\begin{center}
\begin{tabular}{@{ }l@{ }l@{ }c@{ }c@{ }c@{ }} 
\hline
& Parameters& WASP-77 A  & WASP-101  & WASP-103  \\ 
\hline
Star & V mag & 11.3 &  10.3 & 12.1  \\
& Sp. Type  & G8 &F6 & F8  \\
& M$_s$ (M$_\odot$) &1.002 $\pm$ 0.045 & 1.34 $\pm$ 0.07 & 1.22 $\pm$ 0.0039  \\
& R$_s$ (R$_\odot$) &0.955 $\pm$ 0.015 & 1.29 $\pm$ 0.04 & 1.436$^{+0.031}_{-0.052}$ \\
& T$_{\rm{eff}}$ (K) &  5500 $\pm$ 80 & 6380 $\pm$ 120 & 6110 $\pm$ 100  \\
& $v\,\rm{sin}i_*$ (km/s) & 4 $\pm$ 0.2 &12.4 $\pm$ 0.5 & 8.8 $\pm$ 0.7  \\

Planet & M$_p$ (M$_{\rm Jup}$) & 1.76 $\pm$ 0.06 &0.50 $\pm$ 0.04 & 1.49 $\pm$ 0.088  \\
& R$_p$ (R$_{\rm Jup}$)  & 1.21 $\pm$ 0.02 & 1.41 $\pm$ 0.05 & 1.528 $\pm$ 0.03  \\
& Period (d) & 1.3600309 $\pm$ 0.0000020  & 3.585722 $\pm$ 0.000004 & 0.925542 $\pm$ 0.000019   \\
& $\rm{T_0}$ - 2450000 (d)  & 5870.44977 $\pm$ 0.00014& 6164.6934 $\pm$ 0.0002 & 6459.59957  $\pm$ 0.00075   \\
&a (AU)  &0.0240 $\pm$ 0.00036  &0.0506 $\pm$ 0.0009&  	0.01985 $\pm$  0.00021\\
&e  & 0.  & 0. &0. \\
&i (deg)  &89.4$^{+0.4}_{-0.7}$ &85.0 $\pm$ 0.2&89.7 $\pm$ 0.2 \\
&T$_{\rm{eq}}$ (K) &1876 $\pm$ 80   &1565 $\pm$ 35&2508$^{+75}_{-70}$\\

& References & \citet{max13} & \citet{hel14}    & \citet{gill14}  \\
\end{tabular}


\vspace{0.4cm}
\begin{tabular}{@{ }l@{ }l@{ }c@{ }c@{ }c@{ }} 
\hline
& Parameters& WASP-105 & WASP-120 & WASP-131 \\ 
\hline
Star & V mag & 12.1 & 11 & 10.1 \\
& Sp. Type   & K3 & F5 & G0 \\
& M$_s$ (M$_\odot$)  & 0.89 $\pm$ 0.09 & 1.393 $\pm$ 0.057& 1.06 $\pm$ 0.06 \\
& R$_s$ (R$_\odot$)  & 0.90 $\pm$ 0.03 & 1.87 $\pm$ 0.11  & 1.53 $\pm$ 0.05\\
& T$_{\rm{eff}}$ (K) &  5070 $\pm$ 130  &  6450 $\pm$ 120 & 5950 $\pm$ 100\\
& $v\,\rm{sin}i_*$ (km/s) & 1.7 $\pm$ 1.9 & 15.1 $\pm$ 1.2& 3.0 $\pm$ 0.9 \\

Planet & M$_p$ (M$_{\rm Jup}$) & 1.8 $\pm$ 0.1 & 4.85 $\pm$ 0.21 & 0.27 $\pm$ 0.02 \\
& R$_p$ (R$_{\rm Jup}$)  & 0.96 $\pm$ 0.03 & 1.473 $\pm$ 0.096 & 1.22 $\pm$ 0.05 \\
& Period (d) &  7.87288 $\pm$ 0.00001  & 3.6112706 $\pm$ 0.0000043 & 5.322023 $\pm$ 0.000005 \\
& $\rm{T_0}$ - 2450000 (d)  & 6600.0765  $\pm$ 0.0002  &  6779.4352 $\pm$ 0.00044 & 6919.8236 $\pm$ 0.0004 \\

&a (AU)  & 0.075 $\pm$ 0.003& 0.0514 $\pm$0.0007 & 0.0607 $\pm$0.0009\\
&e  &0.& 	0.057$^{+0.022}_{-0.018}$ & 	0. \\
&i (deg)  &89.7 $\pm$ 0.2& 	82.54 $\pm$ 0.78& 	85.0 $\pm$ 0.3 \\
&T$_{\rm{eq}}$ (K) &846 $\pm$ 20& 	1876 $\pm$ 70& 	1460 $\pm$ 30 \\

& References  & \citet{and17} & \citet{tur16} & \citet{hel17} \\
\hline 
\end{tabular}

\end{center}
\end{table*}

\subsection{Rossiter–McLaughlin effect}
\begin{figure}[h]
\includegraphics[width=0.45\textwidth]{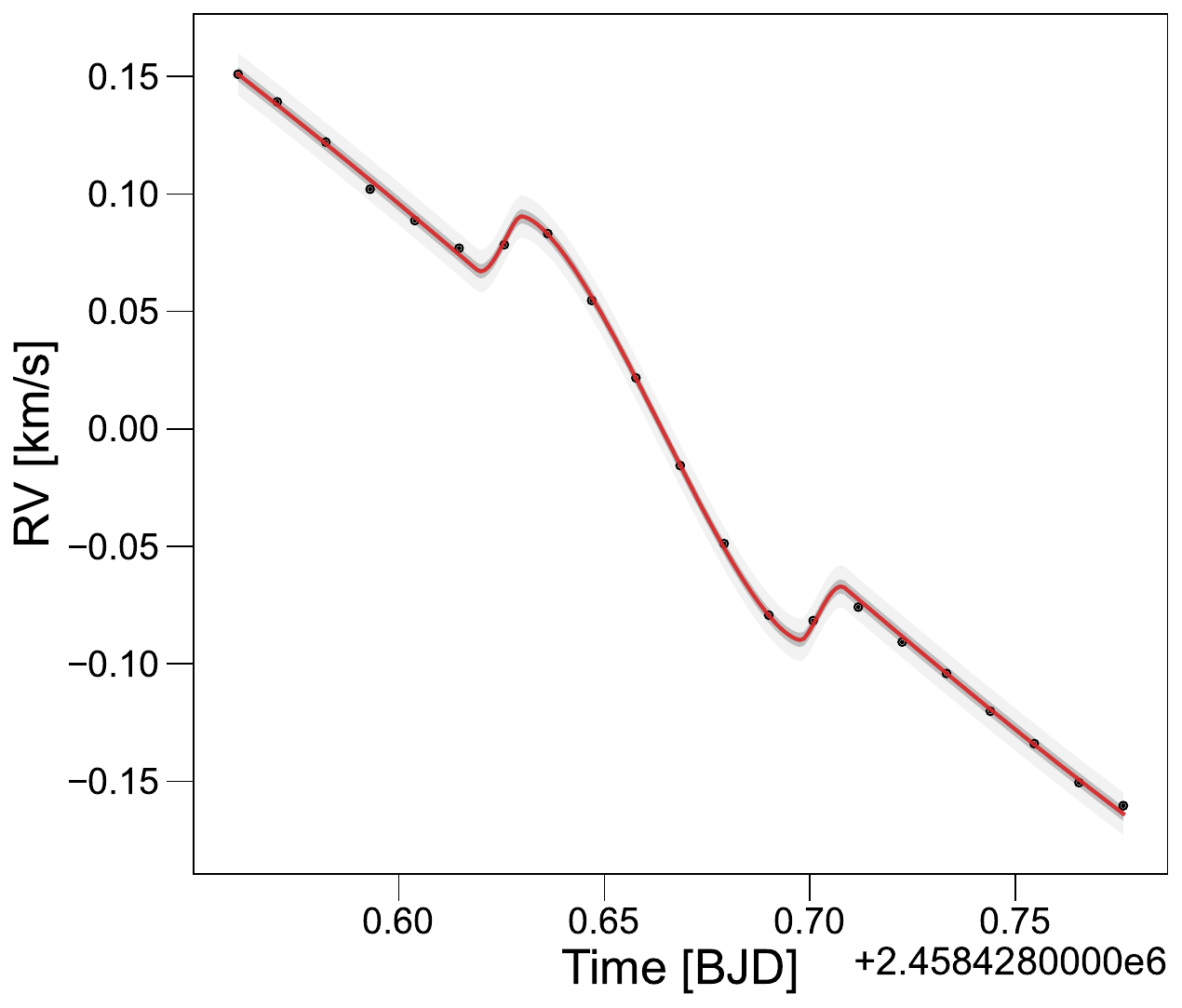}
\caption{The Rossiter-Mclaughlin effect of WASP-77 Ab observed with HARPS. The observed data points (black) are shown with their errorbars, which in this case are smaller than the symbol size. The systemic velocity was removed for better visibility. The red line shows the best fitting model to the data, together with 1-$\sigma$ (dark grey) and 3-$\sigma$ (light grey) confidence intervals.}
\label{f:77}
\end{figure}

\begin{figure}[h]
\includegraphics[width=0.45\textwidth]{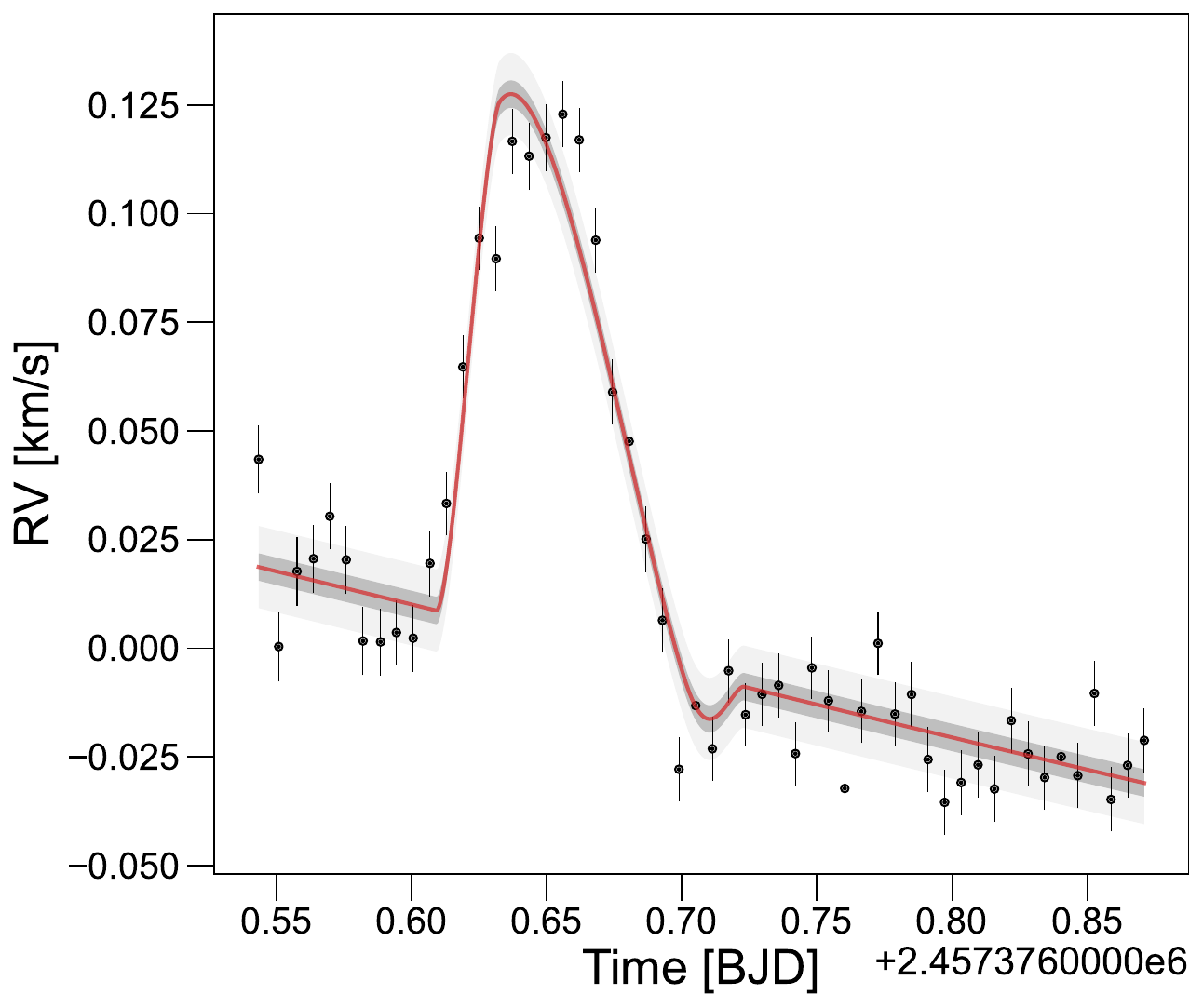}
\caption{Same as Fig.\ref{f:77} for WASP-101b.}
\label{f:101}
\end{figure}

\begin{figure}[h]
\includegraphics[width=0.45\textwidth]{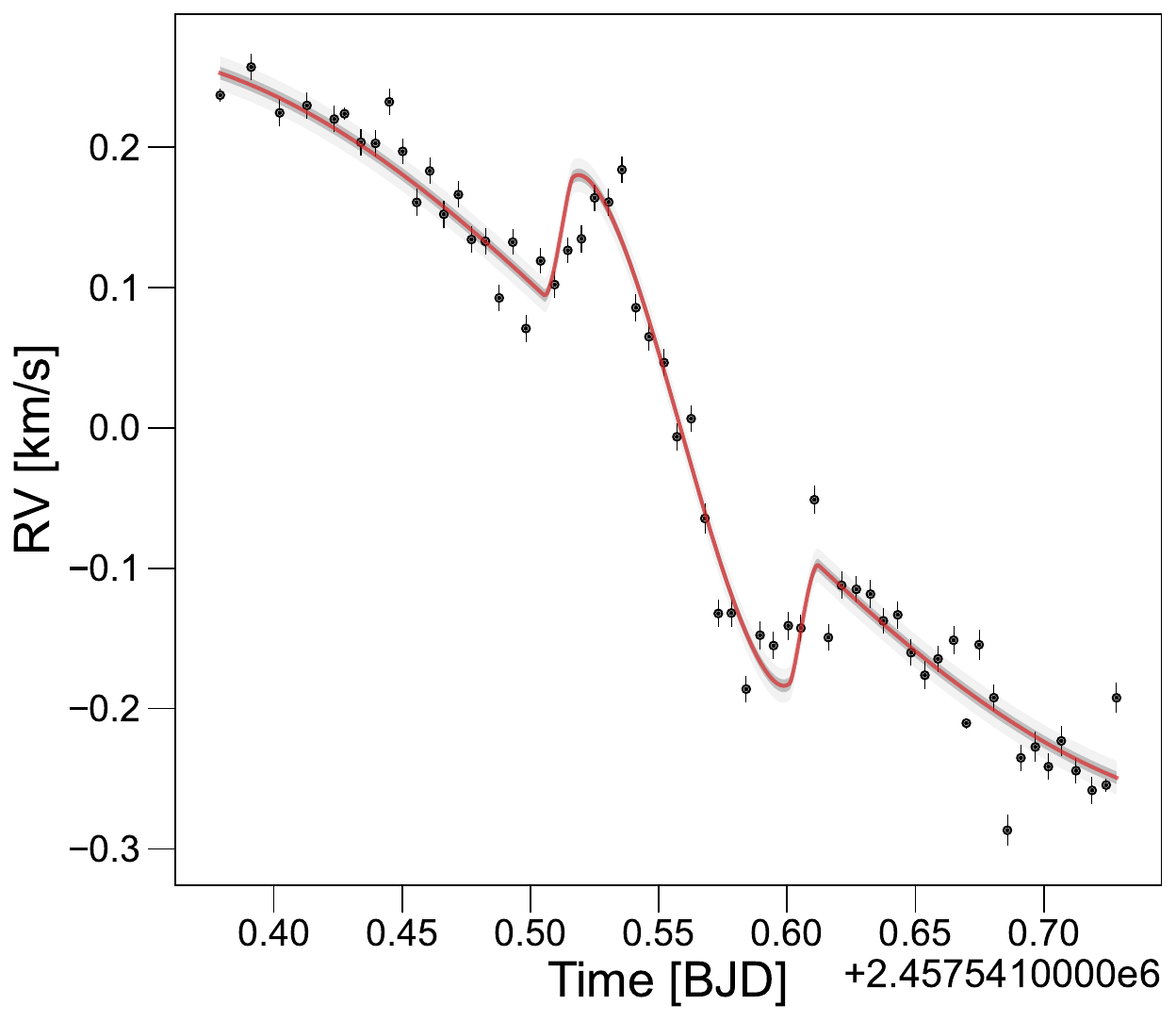}
\caption{Same as Fig.\ref{f:77} for WASP-103b.}
\label{f:103}
\end{figure}

\begin{figure}[h]
\includegraphics[width=0.45\textwidth]{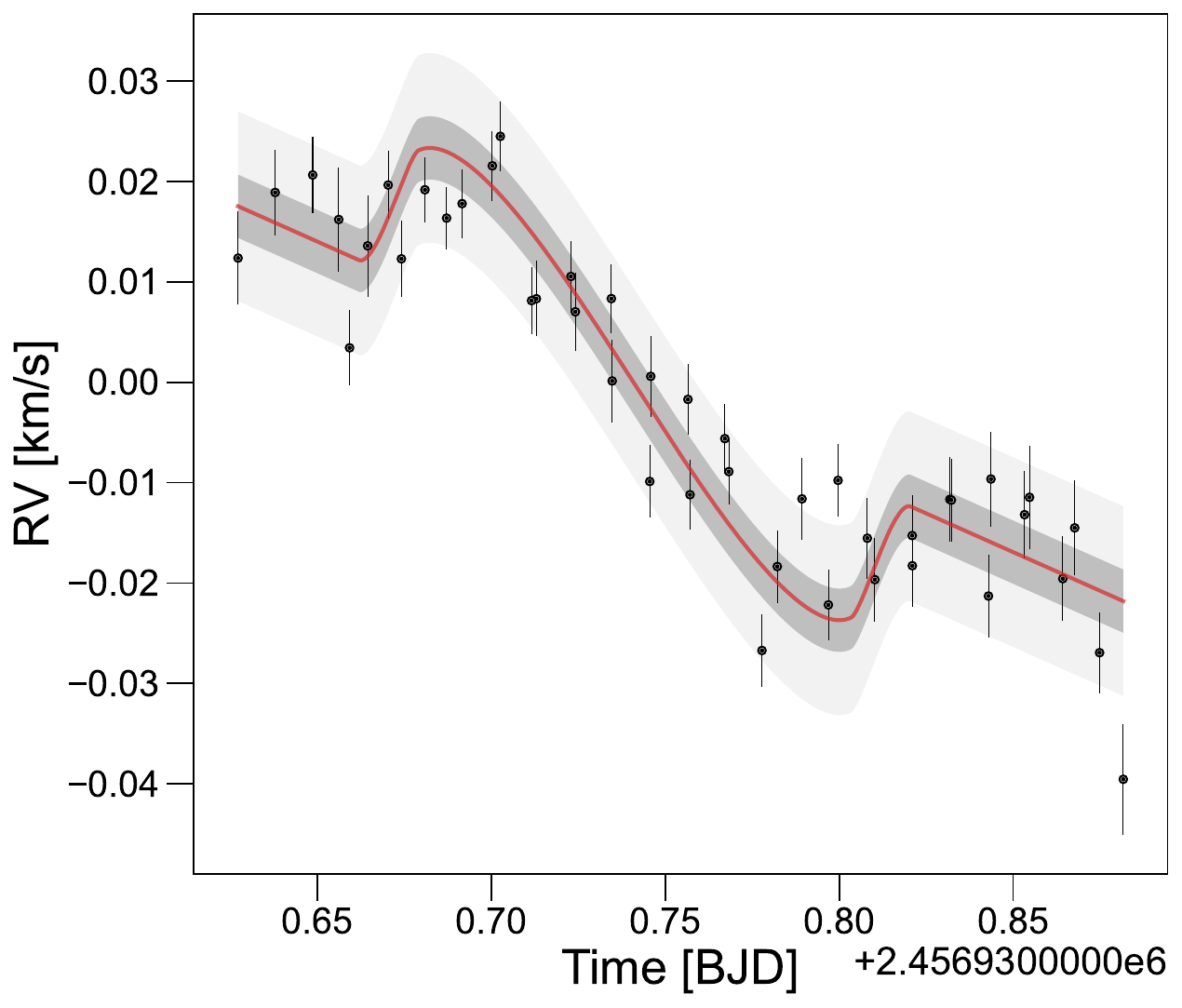}
\caption{Same as Fig.\ref{f:77} for  WASP-105b.}
\label{f:105}
\end{figure}

\begin{figure}[h]
\includegraphics[width=0.45\textwidth]{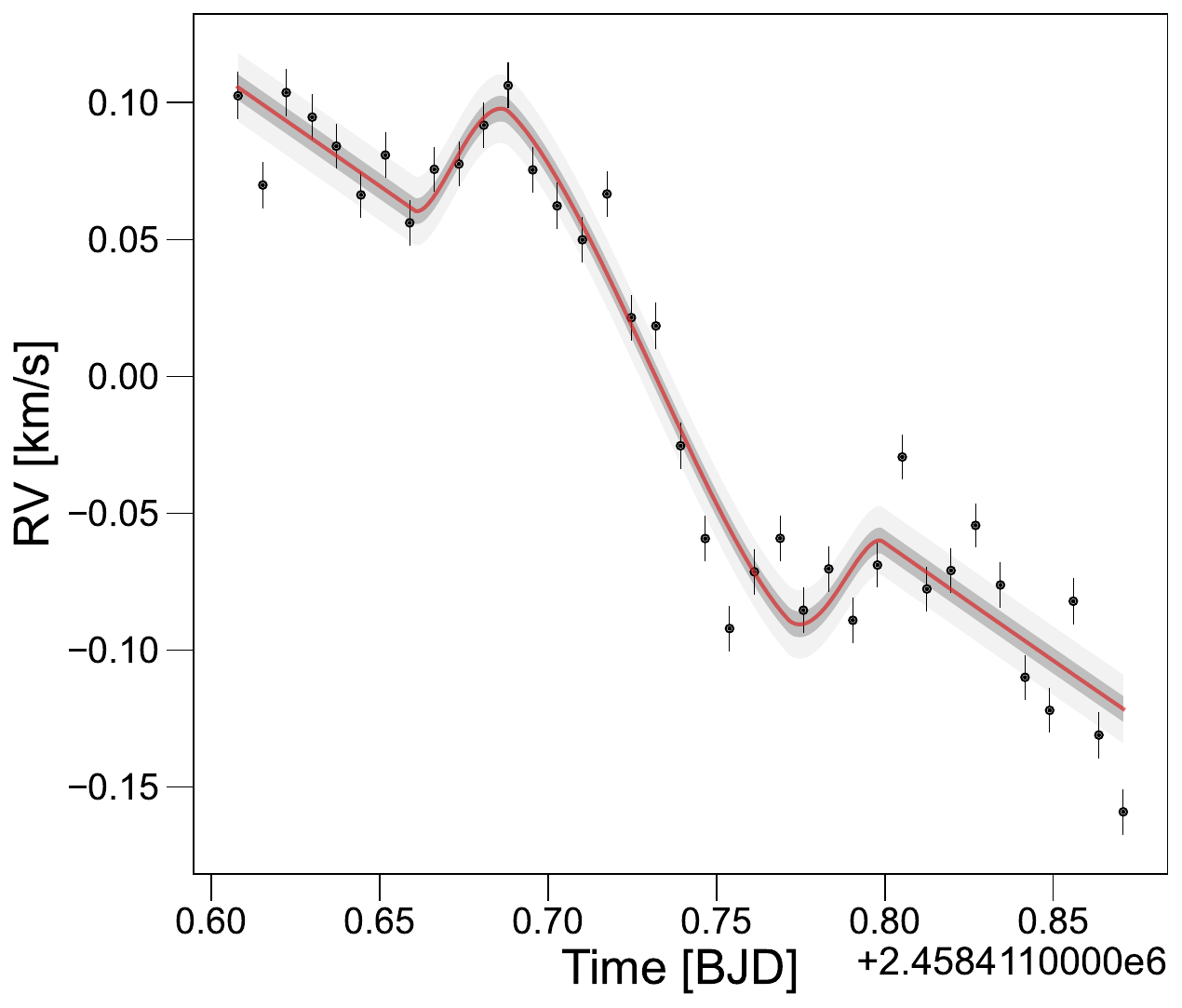}
\caption{Same as Fig.\ref{f:77} for WASP-120b.}
\label{f:120}
\end{figure}

\begin{figure}[h]
\includegraphics[width=0.45\textwidth]{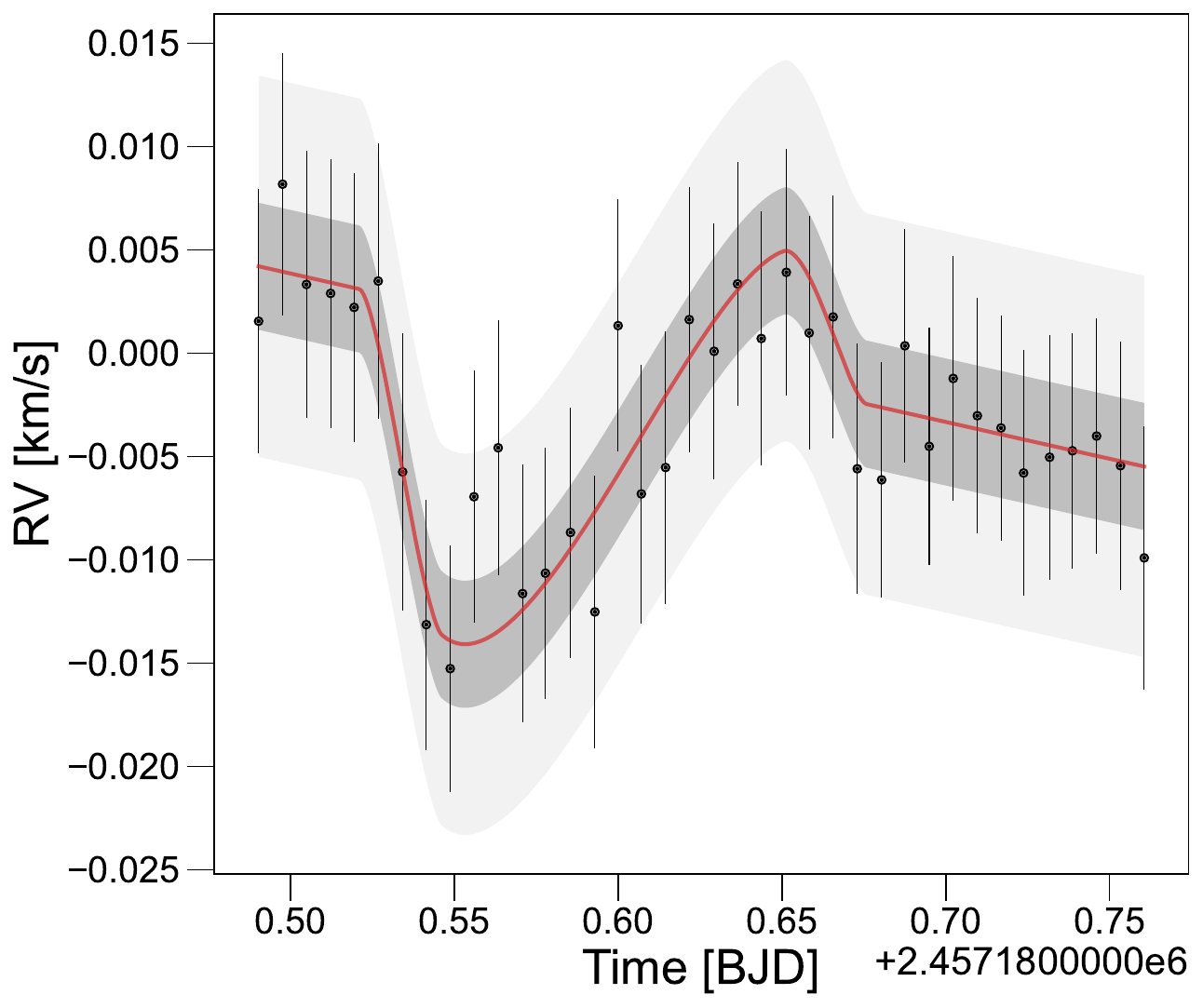}
\caption{Same as Fig.\ref{f:77} for WASP-131b.}
\label{f:131}
\end{figure}

The radial velocities measured during the transit show the anomaly produced by the R-M effect. 
We have retrieved the radial velocity measurements that are obtained from the HARPS and HARPS-N DRS pipeline together with their uncertainties. To measure the projected stellar obliquity of the system ($\lambda$) we fit the radial velocity data with a subroutine from the \textsc{Radvel} package \citep{ful18}  that assumes a Keplerian orbit and the R-M effect formulation of \textsc{ARoMEpy}\footnote{\url{https://github.com/esedagha/ARoMEpy}} \citep{seda23}, which is a python translation of the \textsc{ARoME} code \citep{bou13}. We have used the R-M effect function defined for radial velocities determined through the cross-correlation technique in our code. We fix the radial velocity semi-amplitude $K$ to the literature value and allow the systemic velocity $\gamma$ to be a free parameter.
In the R-M effect model, we fix the following parameters to those reported in the literature: orbital period $P$, planet-to-star radius ratio ${R_p}/{R_s}$ and the eccentricity $e$. The parameter $\sigma$, which is the width of the CCF (FWHM of a Gaussian fit to the CCF), is measured on the data and fixed. Furthermore, we use the ExoCTK\footnote{\url{https://github.com/ExoCTK/exoctk}}  tool to compute the quadratic limb-darkening coefficients in the wavelength range of the HARPS and HARPS-N instruments (380-690 nm). We allow the following parameters to be free during the fitting procedure within the priors from photometry: central transit time $T_C$, projected stellar rotational velocity $v\,\sin i_*$, sky-projected angle between stellar rotation axis and normal of the orbital plane $\lambda$,
orbital inclination $i$, and the scaled semi-major axis ${a}/{R_s}$.
To obtain the best fitting values of the parameters we employed 3 independent Markov Chain Monte Carlo (MCMC) simulations each with 250\,000 steps, burning the first 50\,000.
Using this setup we obtain our results and present them in Sect.~\ref{sec:resrm} and in Figs.~\ref{f:77} to \ref{f:131}.

\subsection{Transmission spectroscopy}
We performed transmission spectroscopy of our targets, following the methodology performed by \citet{wyt15} and \citet{zak19}.
First, to correct for telluric emission lines we have used the fiber B which was aimed at the adjacent sky and used the {\tt Molecfit} software \citep{sme15,kau15}. {\tt Molecfit} uses synthetic modeling of the Earth's atmospheric spectrum with a line-by-line approach. We carefully select nine telluric wavelength regions in the range of 589 to 651 nm. These regions contain unblended telluric lines produced by $\rm{H_2O}$ and $\rm{O_2}$. The best-fit {\tt Molecfit} solution was applied to the whole HARPS spectra.
As the next step, we shift all the spectra to the stellar rest frame. We have performed a wavelength cut to isolate the region of interest and performed flux normalization for the spectra. Normalization was done in IRAF \citep{tod87} using the task {\tt continuum} and on the regions of the spectra without strong lines.
Subsequently, we have averaged all out-of-transit spectra to create a stellar template. We divided each spectrum during transit by this template to obtain individual transmission spectra. Such spectra need to be corrected for the planetary motion of the planet. These corrected residuals were summed up and normalized to unity. Finally, unity was subtracted to obtain the final transmission spectrum $\tilde{\mathcal{R}}(\lambda)$:

\small
\begin{center}
\begin{equation} \label{trans}
\tilde{\mathcal{R}}(\lambda)=\sum_{\rm{in}} \left. \frac{\tilde{f}_{\rm{in}}(\lambda,t_{\rm{in}})}{\tilde{F}_{\rm{out}}(\lambda)}\right|_{\rm{Planet\ RV\ shift}} -1
\end{equation}
\end{center}
\normalsize

In principle, stellar activity might hinder our ability to retrieve any planetary signal \citep[e.g.,][]{barn17}. To assess the impact of stellar activity on the transmission spectra, we have repeated our methodology on several control lines (Mg {\sc i} and Ca {\sc i}). We retrieve a flat spectrum consistent with a negligible effect of the stellar activity.

\section{Results} \label{sec:res}
\subsection{Projected stellar obliquity measurement}
\label{sec:resrm}

We have measured the projected stellar obliquity ($\lambda$) of four targets for the first time: WASP-77~Ab, WASP-101b, WASP-105b, and WASP-120b. WASP-103b has been previously studied by \citet{add16} using the UCLES spectrograph. WASP-131b had its spin-orbit alignment measured by \citet{doy23} using ESPRESSO during the advanced stages of preparing this manuscript.

\textbf{WASP-77 Ab} is an inflated hot Jupiter orbiting a G8 host star on a short 1.4\,day orbit. The host star has a fainter companion at a distance of 3 arcsec \citep{max13}. We infer an aligned orbit of WASP-77 Ab with $\lambda =-8^{+19}_{-18}$\,deg.

\textbf{WASP-101b} is a highly inflated hot Jupiter orbiting a F6 host star on a 3.5\,day orbit.
We infer a misaligned orbit of WASP-101b with $\lambda$ =34 $\pm$ 3\,deg.

\textbf{WASP-103b} is an ultra-hot Jupiter orbiting an F8 host star on a 0.92\,day orbit. \citet{woll15} discovered a nearby source that \citet{south16} characterized as a $0.72 \pm 0.08$ M$_\odot$ star in the WASP-103 system. The system has been frequently observed \citep[e.g.,][]{mac18,del18,pat20} for possible period changes. Recent CHEOPS observations found hint of an orbital period increase due to tidal decay \citep{bar22}.
We infer an aligned orbit of WASP-103b with $\lambda =2^{+35}_{-36}$\,deg. The derived projected obliquity is in agreement with the results by \citet{add16} who obtained $\lambda =3 \pm 33$\,deg.

\textbf{WASP-105b} is a warm Jupiter-sized planet. It orbits its K2 host star on a 7.8\,day orbit. We infer an aligned orbit with $\lambda =-14^{+28}_{-24}$\,deg.

\textbf{WASP-120b} is a massive hot Jupiter. It orbits its F5 host star on a 3.6\,day, slightly eccentric orbit ($e=0.05$). \citet{bohn20} found a possible binary companion, but further astrometric data are needed to confirm the tertiary nature of the system. We find that the orbit of WASP-120b is well aligned, with $\lambda$ =-2 $\pm$ 4\,deg.

\textbf{WASP-131b} is a highly inflated hot Saturn. It orbits its G0 host star on a 5.3\,day orbit. WASP-131 has a low mass stellar companion \citep{bohn20,south20}.
We infer a highly misaligned orbit of WASP-131b, with $\lambda$ =161 $\pm$ 5\,deg. Our results based on  HARPS data are in excellent agreement with the results from ESPRESSO, $\lambda =162.4^{+1.3}_{-1.2}$\,deg \citep{doy23}.

The results of our fits are shown in Figs. \ref{f:77}--\ref{f:131}, while the MCMC results are shown in Figs.~\ref{m77} to \ref{m131} of the Appendix. The derived values from the MCMC analysis are displayed in Table~\ref{tab:res3}. 

The values of $v\,\sin{i_*}$ that we derived in our R-M effect analysis are generally in agreement with the ones reported in the literature from spectral synthesis.
A discrepancy is, however, observed for WASP-77 A. Hence, we have measured the $v\,\sin{i_*}$ from our HARPS stacked spectra, using {\tt iSpec} \citep{cua14}. We obtained a value of $v\,\sin{i_*}$ = 2.6 $\pm $ 0.4 km/s, in better agreement with the value obtained from our R-M effect analysis in Table~\ref{tab:res3}.

\begin{table*}[htbp]
\caption{MCMC analysis results.}
\label{tab:res3}
\begin{center}
\begin{tabular}{l l | c c | c c | c c} 

\hline
& Parameter & Prior & WASP-77 A & Prior & WASP-101 & Prior & WASP-103 \\ 
\hline
  \vspace{0.05 cm}
 & $T_c$ - 2450000 [d] 
 & $\mathcal{N}(T_0, 0.006)$ & 8428.664$^{+0.001}_{-0.001}$ 
 & $\mathcal{N}(T_0, 0.003)$ & 7376.666 $\pm$ 0.002
 & $\mathcal{N}(T_0, 0.006)$ & 7540.632$^{+0.002}_{-0.001}$ \\
  \vspace{0.05 cm}
 & $\lambda$ [deg] 
 & $\mathcal{U}(-180, 180)$ &  -8$^{+19}_{-18}$
 & $\mathcal{U}(-180, 180)$ & 34 $\pm$ 3
 & $\mathcal{U}(-180, 180)$ & 2$^{+35}_{-36}$ \\
 \vspace{0.05 cm}
 & $v\,\sin{i_*}$ [km/s] 
 & $\mathcal{U}(0, 10)$ & 2.0$^{+0.2}_{-0.1}$ \ 
 & $\mathcal{U}(0, 15)$ &  8.9 $\pm$ 0.6 
 & $\mathcal{U}(0, 10)$ & 7$^{+2}_{-1}$ \\
  \vspace{0.05 cm}
 & $a/R_s$ 
 & $\mathcal{N}(5.4, 0.2)$ & 5.53  $\pm$ 0.05
 & $\mathcal{N}(8.3, 0.3)$ & 8.3 $\pm$ 0.2
 & $\mathcal{N}(2.99, 0.05)$ & 2.99$^{+0.02}_{-0.02}$  \\
  \vspace{0.05 cm}
 & Inc. [deg] 
 & $\mathcal{N}(89.4, 0.5)$ & 89.52$^{+0.36}_{-0.40}$
 & $\mathcal{N}(85.0, 0.2)$ &  84.88 $^{+0.39}_{-0.34}$
 & $\mathcal{N}(89.7, 0.2)$ & 89.43$^{+0.4}_{-0.4}$  \\
  \vspace{0.05 cm}
 & $\gamma$ [km/s] 
 & $\mathcal{N}(1.7, 0.2)$ & 1.729 $\pm 0.001$
 & $\mathcal{N}(42.6, 0.2)$ & 42.620 $\pm 0.002$
 & $\mathcal{N}(-42.2, -0.2)$ & -41.878 $\pm 0.004$ \\
\hline
& Parameter & Prior & WASP-105 & Prior & WASP-120 & Prior & WASP-131 \\ 
\hline
 \vspace{0.05 cm}
 & $T_c$ - 2450000 [d] 
 & $\mathcal{N}(T_0\pm 0.006)$ & 6930.740$^{+0.004}_{-0.004}$ 
 & $\mathcal{N}(T_0\pm 0.006)$ & 8411.730$^{+0.003}_{-0.003}$
 & $\mathcal{N}(T_0\pm 0.006)$ &  7180.598$\pm 0.003$\\
 \vspace{0.05 cm}
 & $\lambda$ [deg] 
 & $\mathcal{U}(-180, 180)$ & -14 $^{+28}_{-24}$ 
 & $\mathcal{U}(-180, 180)$ &  -2 $\pm$ 4
 & $\mathcal{U}(-180, 180)$ & 161 $\pm$ 5 \\
\vspace{0.05 cm}
 & $v\,\sin{i_*}$ [km/s] 
 & $\mathcal{U}(0, 2)$ & 0.8$^{+0.2}_{-0.2}$ 
 & $\mathcal{U}(0, 18)$ & 14.5$^{+2.2}_{-2.6}$
 & $\mathcal{U}(0, 5)$ & 2.17$^{+0.33}_{-0.31}$ \\
 \vspace{0.05 cm}
 & $a/R_s$ 
 & $\mathcal{N}(17.2,0.3)$ & 17.2$^{+0.3}_{-0.3}$ 
 & $\mathcal{N}(5.9, 0.2)$ & 5.91$^{+0.19}_{-0.19}$ 
 & $\mathcal{N}(8.5, 0.2)$ &  8.52$^{+0.20}_{-0.20}$ \\
 \vspace{0.05 cm}
 & Inc. [deg] 
 & $\mathcal{N}(89.2, 0.2)$ & 89.20$^{+0.21}_{-0.20}$ 
 & $\mathcal{N}(82.54, 0.78)$ & 82.15$^{+0.48}_{-0.44}$
 & $\mathcal{N}(85.0, 0.3)$ & 85.01$^{+0.39}_{-0.34}$ \\
 \vspace{0.05 cm}
 & $\gamma$ [km/s] 
 & $\mathcal{N}(24.6, 0.2)$ & -24.684$\pm 0.001$
 & $\mathcal{N}(19.8, 0.2)$ & 19.827$\pm 0.004$
 & $\mathcal{N}(-19.6, 0.2)$ & -19.638$\pm 0.001$ \\
\hline
\end{tabular}
\end{center}
\end{table*}

\begin{table}
\caption{Transmission (TSM) and emission (ESM) spectroscopy metrics for our sample, as defined in \citet{kemp18}.}
\vspace{-.4cm}
\label{tab:met}
\begin{center}
\begin{tabular}{lcccccc}
\hline
WASP-& 77b &101b &103b& 105b & 120b & 131b  \\ 
\hline
TSM& 186&365&89 & 21 & 17 & 356 \\
ESM&413&147&133 & 36 & 68 & 203\\
\hline 
\end{tabular}

\end{center}
\end{table}

\subsection{Transmission spectroscopy}
\label{resTS}
\begin{figure}[h]
\includegraphics[width=0.5\textwidth]{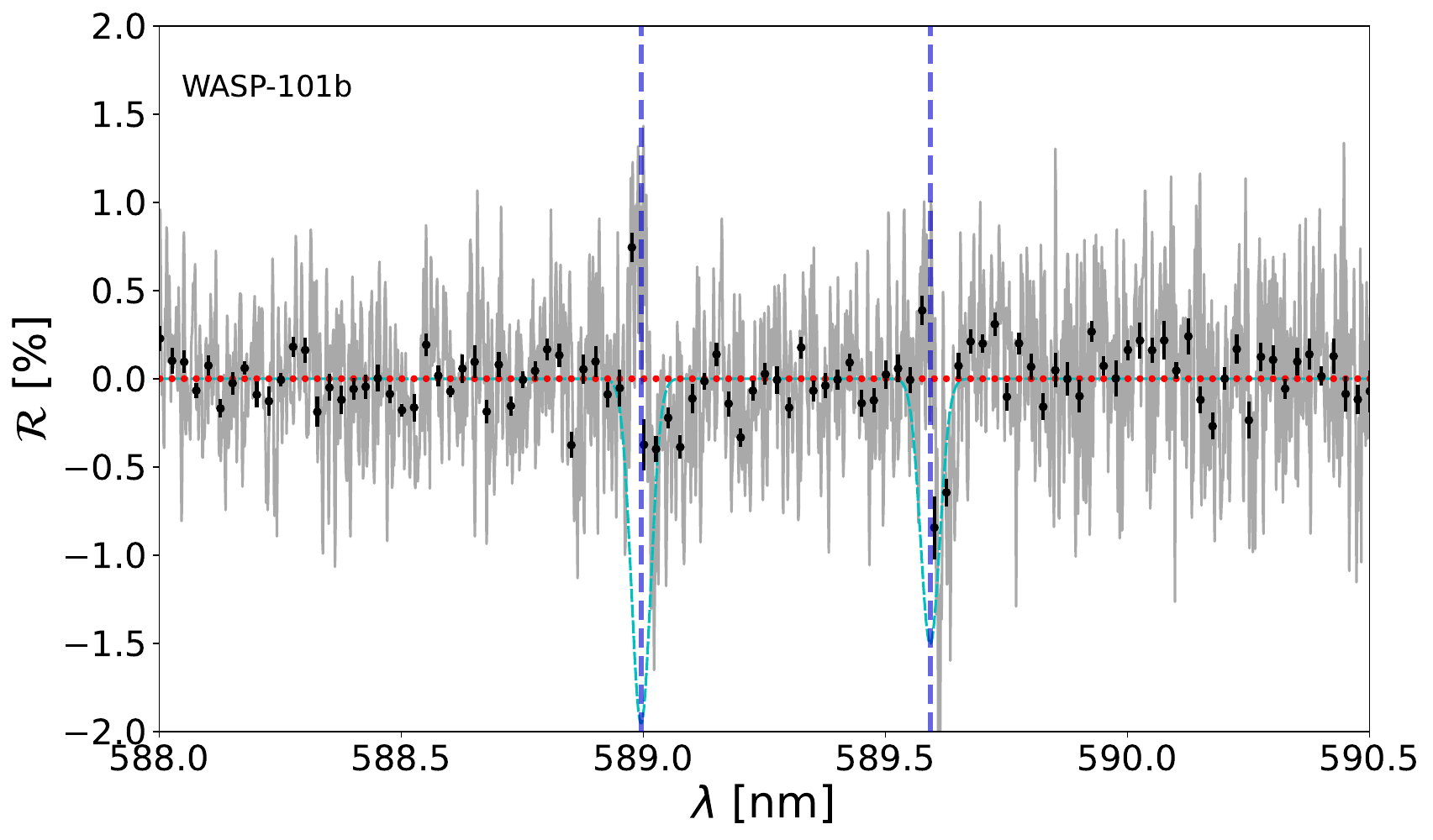}
\caption{Transmission spectrum for WASP-101b showing a non-detection of any absorbers. The blue dashed lines indicate the position of the Na D2 and D1 lines. We show in cyan a simulated spectrum corresponding to a clear atmosphere model. In red we show a featureless spectrum. The latter model is preferred by a $\chi^2$ test.}
\label{W101tsna}
\end{figure}

\begin{figure}[h]
\includegraphics[width=0.5\textwidth]{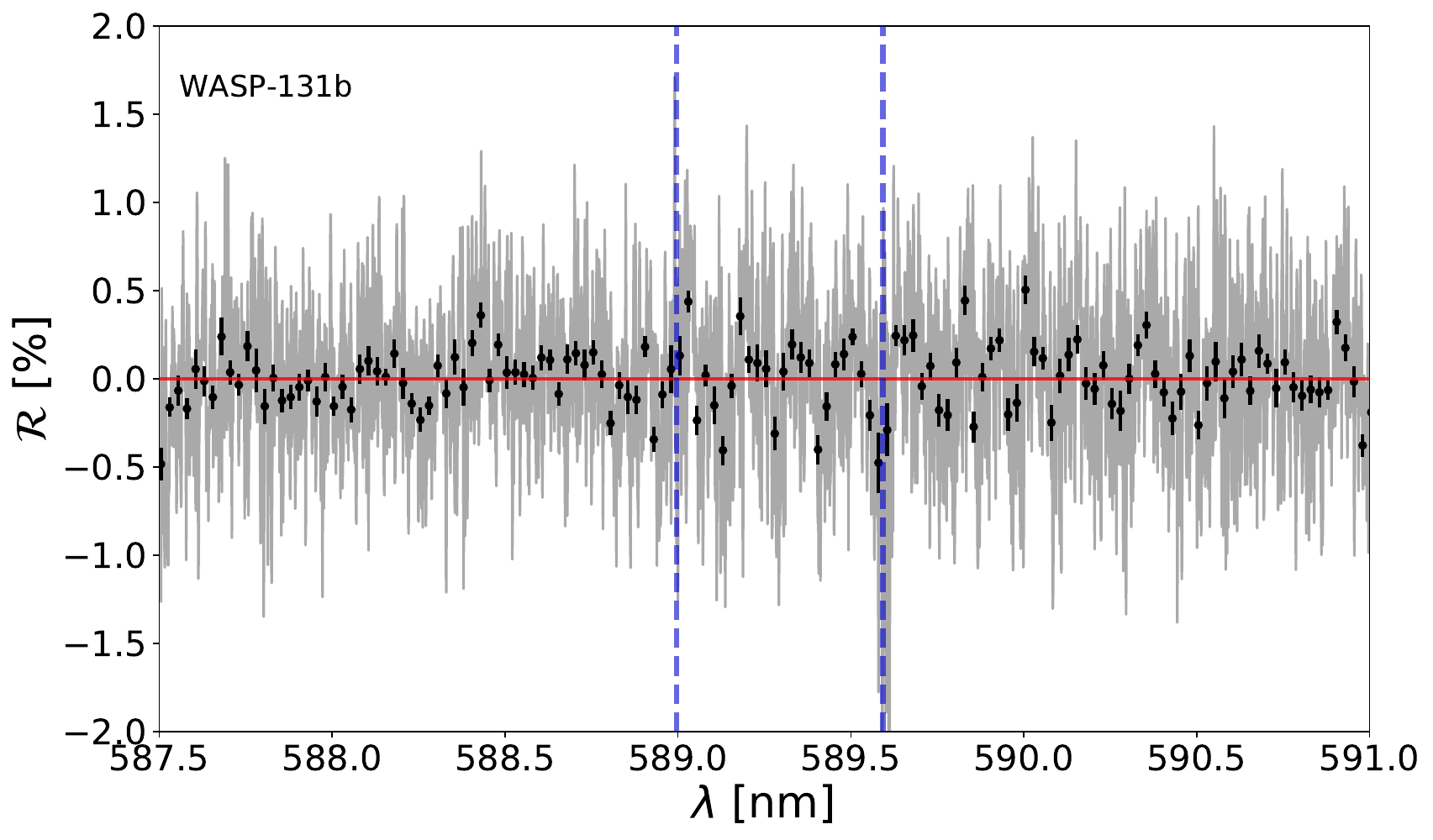}
\caption{Transmission spectrum for WASP-131b showing a non-detection of any absorbers. The blue dashed lines indicate the position of the Na D2 and D1 lines.}
\label{W131tsna}
\end{figure}

With the same data sets that were used to measure the projected stellar obliquity we also performed transmission spectroscopy to search for strong optical absorbers such as sodium (589\,nm doublet) and hydrogen (Balmer alpha). The most amenable targets are WASP-101b and WASP-131b, thanks to their high transmission spectroscopy metric, which rates the suitability of the target for atmospheric characterization \citep[][]{kemp18} -- see Table~\ref{tab:met}. For all the targets, we derive featureless spectra and show in Figs. \ref{W101tsna} and \ref{W131tsna} the sodium doublet regions for WASP-101b and WASP-131b around the sodium doublet. The results for WASP-101b are in agreement with the previous studies using the HST in the near-infrared (NIR) and optical region \citep{wak17101,rat23} that found featureless spectra and attributed them to the presence of thick clouds. While in Fig. \ref{W101tsna} the sodium D1 appears to have two points in the absorption, we do not claim a detection, especially as the sodium D2 line is predicted to have a more profound line profile \citep{geb20} that we do not detect in our spectra. We have compared the observed spectrum with a clear atmosphere model and a featureless spectrum using a $\chi^2$ test and the featureless model is preferred.

WASP-103b is an accessible target for the investigation of Balmer-driven mass loss due to its proximity to the host star \citep{gar19}. We derive a featureless spectrum around the H-alpha region. We provide the first transmission spectrum of WASP-131b which is expected to have a large atmospheric signal. The derived spectrum shows no absorption features and possibly hints at either, the presence of clouds, Rayleigh scattering muting the strength of the absorption features, and/or processes depleting sodium \citep{iro05}.

\section{Discussion}

\subsection{Misaligned planets}
\begin{figure*}[h]
\centering
\includegraphics[width=0.7\textwidth]{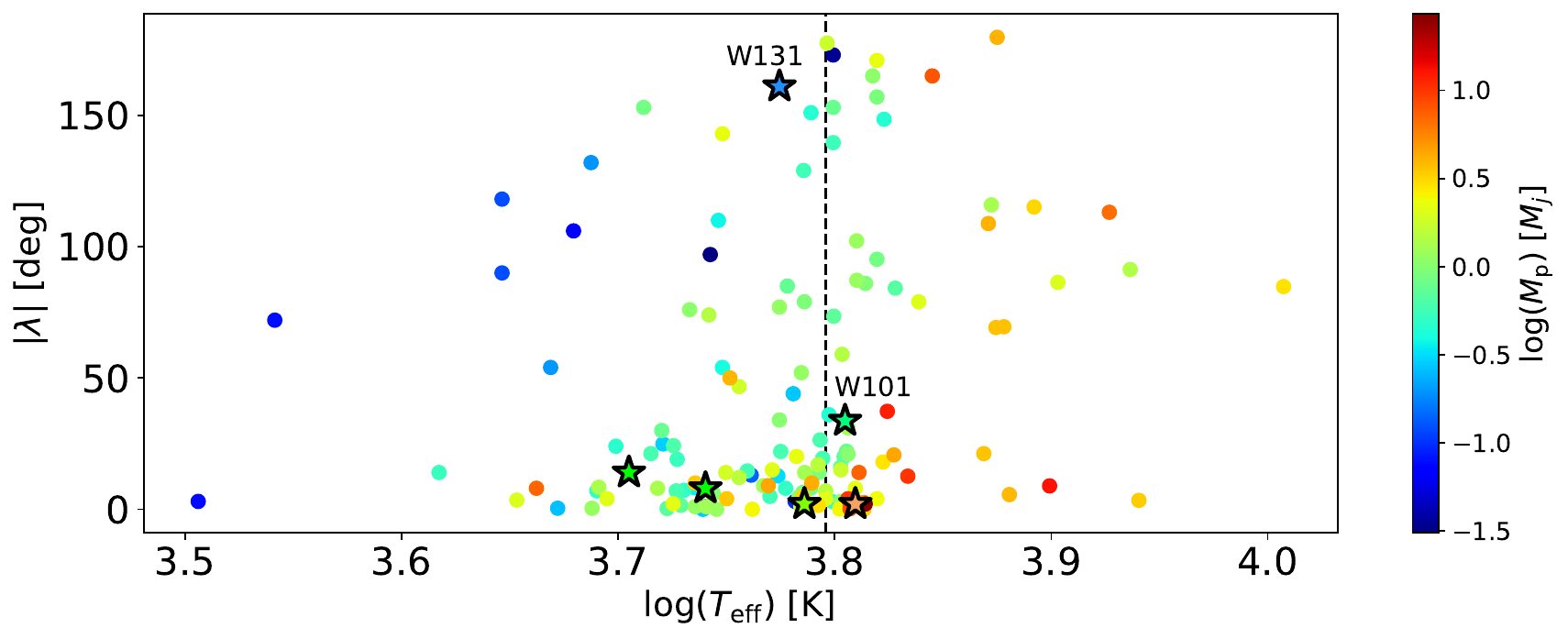}
\caption{Projected obliquity versus stellar effective temperature.
Our targets are plotted with stars. 
The dashed vertical black line shows the position of the Kraft break.}
\label{lamteff}
\end{figure*}

We compare our results to the known sample\footnote{Retrieved from \url{https://exoplanetarchive.ipac.caltech.edu/}} of exoplanets with measured spin-orbit alignment (Fig. \ref{lamteff}). WASP-131b resides in the less populated region of highly-misaligned planets with host star temperature lower than 6250 K. This temperature corresponds to the Kraft break, which separates stars with deep convective envelopes and efficient magnetic dynamos from those without \citep{maed09,ava22}. As noted in \citet{alb22}, most of the planets that are on retrograde orbits around cooler stars are planets with Neptunian masses. WASP-131b is one of the most misaligned Saturn-mass planets known and is an especially intriguing target for migration theories as it is known that its host has a stellar companion on wide orbit \citep{bohn20}. The Kozai-Lidov mechanism \citep{fab07} can put a planet on a retrograde orbit in the presence of a wide stellar companion. As the orbit has already been circularized through tidal interactions with the host star, we cannot confirm this scenario. However, future atmospheric chemistry studies may be able to provide clues about the position in the disc where the planet formed \citep{mad14}.

WASP-101b with its projected obliquity of 34 degrees joins a less populated region of misaligned planets (Fig. \ref{lamteff}). A trend \citep{al21,att23} suggests that misaligned planets do not equally cover the full range of obliquities but they rather show a preference for polar orbits ($\lambda$$\sim$90 deg). However, the mechanisms responsible for this trend are still under debate \citep{lai12,pet20,sie23}. Furthermore, \citet{dong23} reported that misaligned systems have nearly isotropic stellar obliquities with no strong clustering near 90 degrees. These contradictory findings warrant a further investigation of the distribution of misaligned planets.
The other targets join a well-populated parameter space of aligned gas giant exoplanets.

We tried to assess whether misaligned Jupiter-mass planets around hot stars belong to a different population of exoplanets. To this end, we performed\footnote{We used the 2D KS test Python implementation available at \url{https://github.com/syrte/ndtest}.} a two-dimensional KS test \citep{pea83} to study the parameter space  T$_{\rm{eff}}$-M$_{\rm{p}}$ and compared the misaligned population ($\lambda$$\geq$20 deg) to the whole gas giant population. The test reveal that, with a $p$-value of $1\times10^{-11}$, we can reject the null hypothesis that both groups originate from the same planetary population.

 \subsection{Stellar rotation and true obliquity $\Psi$ }
Photometric light curves can sometimes be used to derive the stellar rotation periods \citep[e.g.,][]{ska22} due to modulations imprinted by the presence of active regions on the spinning star -- a key ingredient to derive the stellar inclination $i_*$ and thus the true obliquity. We have tried to infer the stellar rotation periods using TESS \citep{ri15} and ASAS-SN \textit{g} band \citep{ko17} light curves, using a Lomb-Scargle periodogram \citep{lomb76,sca82}. We were able to detect a periodic modulation of WASP-77 A with a period of $16.2 \pm 0.25$ days in the ASAS-SN data (FAP=$1\times10^{-5}$, Fig. \ref{fig:per}). This is in agreement with a literature value for rotational modulation for WASP-77 A detected by \citet{max13} who reported a value of $15.4 \pm 0.4$ days in the WASP light curve. For other targets, we were not able to detect any signal with high significance.
Using the values above we infer $i_*= 42 \pm 12$\,deg for WASP-77\,A\footnote{We used the obtained $v\,\rm{sin}i_*$ from the R-M effect as the value from spectroscopy is a combination of several broadening mechanisms \citep[e.g.,][]{tri15} and is therefore not a good indicator of the rotation period of the star.}. Combining the $\lambda$ and $i_*$ measurements we are able to infer the true stellar obliquity $\Psi$ using the spherical law of cosines, $\rm{cos}\Psi =\rm{sin}i_*\,\rm{sin}i\,\rm{cos}|\lambda|+\rm{cos}i_*\,\rm{cos}i$. We obtain $\Psi = $48$^{+22}_{-21}$\,degrees\footnote{We note that given the errorbars the system is still aligned within just over 2 sigmas.}.

\begin{figure*}[h]
\centering
\includegraphics[width=0.7\textwidth]{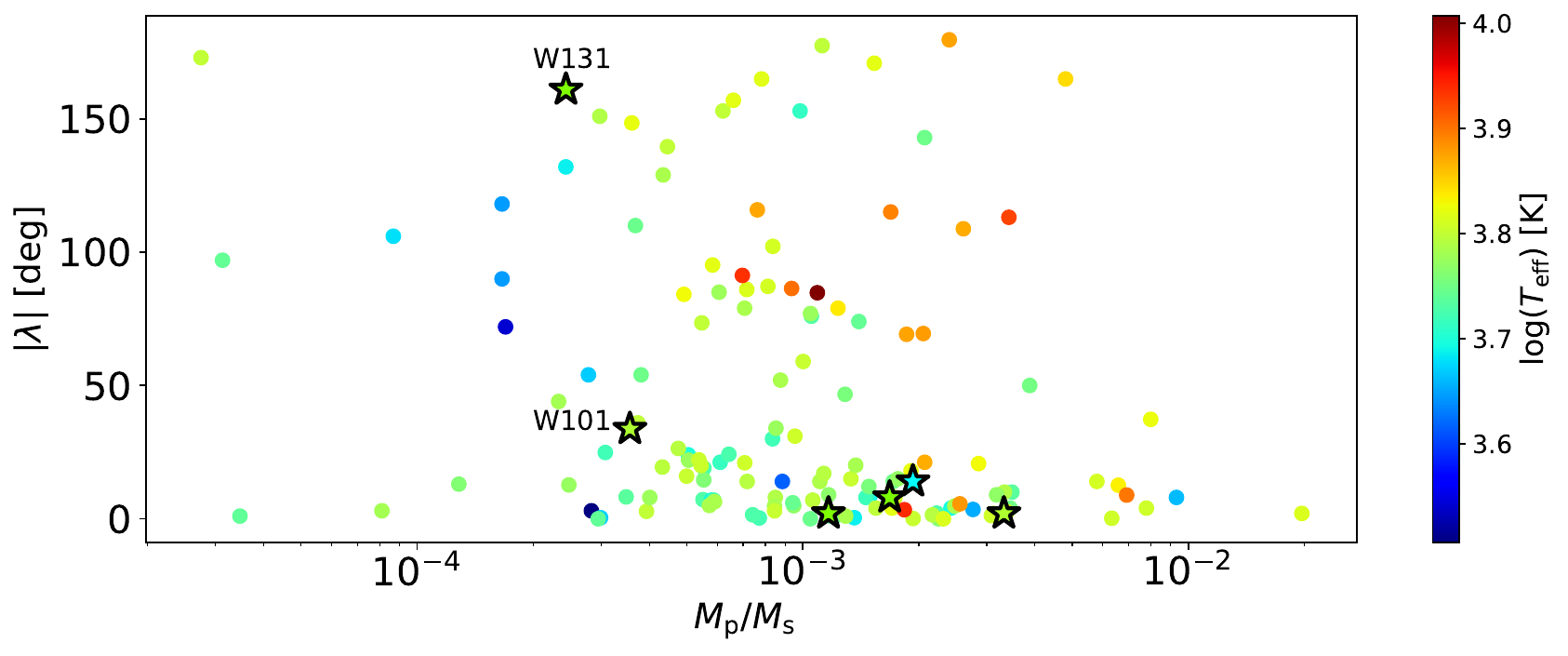}
\caption{Projected obliquity versus the ratio of the planetary to the stellar mass.
Our targets are plotted with stars. 
}
\label{lammass}
\end{figure*}






\subsection{Future atmospheric characterization}
The high-resolution transmission spectroscopy method can probe the upper atmospheres of exoplanets by resolving the deep line cores of strong atomic features. Optical spectra of exoplanets are valuable tools for atmospheric retrievals using broadband infrared data. Such retrievals suffer from severe degeneracies as multiple combinations of parameters can well fit the same data set. These degeneracies can be lifted by constraints provided with optical data \citep{wel19}. The presence of the refractory elements is used for studying the evolution of gas planets \citep{lot21,hands22}. Our results from section \ref{resTS} will aid future missions in obtaining more precise abundances of these gas giants to unravel their history.


In the future, space and airborne missions (e.g., Ariel, \citeauthor{tin18}~\citeyear{tin18}; Twinkle, \citeauthor{edw19}~\citeyear{edw19} and EXCITE, \citeauthor{nag22}~\citeyear{nag22}) will provide a homogeneous sample of exoplanetary atmospheric spectra. ESA's Ariel\footnote{\url{https://arielmission.space/}} mission, 
scheduled for launch in 2029, 
is specifically designed to conduct the first unbiased survey of a statistically significant sample of around 1,000 transiting exoplanets across the visible and near-infrared spectrum. With limited prior information on the targets (mainly limited to their existence, radius, mass, and equilibrium temperature), Ariel's observational strategy is based on four distinct tiers \citep{Edwards2019a, Bocchieri2023}. Tier 3 consists of repeated observations of a select group of benchmark planets (approximately 50-100) around bright stars to characterize their atmospheres in detail.

Knowing the spin-orbit orientation of planetary systems will complement the information obtained by these missions especially in regard to the planetary formation and evolution questions. In the following subsections, we summarize atmospheric prospects for WASP-77Ab, WASP-101b, WASP-103b, and WASP-131b as these planets are the most amenable for future atmospheric studies as suggested by their Transmission Spectroscopy Metric (TSM) and Emission Spectroscopy Metric (ESM), reported in Table~\ref{tab:met}. 

For WASP-77Ab, WASP-101b and WASP-131b, we show the simulated atmospheric transmission spectra as would be obtained by {Ariel}. These spectra were produced by combining forward models from TauREx~3 \citep{2021ApJ...917...37A} with the {Ariel} radiometric simulator, ArielRad \citep{2020ExA....50..303M}, following the same methodology as presented in \citet{2021AJ....162..288M}. That is, for a given planet and atmospheric model, we used TauREx~3 to generate the high-resolution transmission spectrum. Then, we binned the raw spectrum to the Tier-3 spectral grid (R = 20, 100, and 30, in NIRSpec, AIRS-CH0, and AIRS-CH1, respectively) and added the associated noise predicted by ArielRad for each spectral bin. ArielRad is a wrapper of ExoRad~\citep{Mugnai2023} (the instrument-independent version of the radiometric simulator, available on GitHub\footnote{\url{https://github.com/ExObsSim/ExoRad2-public}}) and ArielRad-Payloads, the repository of configuration files for the payload maintained by the Ariel Consortium. For reproducibility, we report the code versions in Table~\ref{tab:ver}. Finally, for WASP-103b, we show predicted JWST MIRI observations. The results for individual planets are presented in Appendix \ref{atmopl}.

\begin{table}[!htbp]
    \caption{Codes used to produce the simulated Ariel spectra.}
    \label{tab:ver}
    \centering
    \begin{tabular}{c@{ }c@{ }}
    \hline
    \bf{Code}     & \bf{Version} \\ 
    \hline
    TauREx~3          & 3.1.1-alpha      \\    
    ArielRad          & 2.4.26           \\
    ExoRad            & 2.1.111          \\
    ArielRad-Payloads & 0.0.17           \\ 
    \hline
    \end{tabular}

\end{table}

\section{Summary}

We analyzed HARPS and HARPS-N archival data of 6 gas giants planets on short orbits around F-G-K dwarfs. We studied the Rossiter-McLaughlin effect and measured the spin orbit alignment for the first time for four planets. We find that WASP-131b is on a highly misaligned orbit and joins a small number of highly misaligned planets around host star with effective temperature below 6250 K. WASP-101b is on a misaligned orbit, while WASP-77 Ab, WASP-103b, WASP-105b and WASP-120b are on aligned orbits. We also derived the true (3-D) obliquity of WASP-77 Ab, $\Psi = $48$^{+22}_{-21}$\,deg. We further performed transmission spectroscopy in order to search for strong atomic absorbers (sodium doublet and H-alpha), but obtained featureless spectra likely due to a cloud deck at high altitudes and/or Rayleigh scattering muting the strength of the features. Finally, we discussed future perspectives for studying these targets with the space missions that will be able to constrain the evolution and migration histories of these planets via atmospheric chemistry studies.

\begin{acknowledgements}
The authors would like to thank the anonymous referee for their insightful report. We would like to also thank Prune C. August for sending us model spectra.
MV, MS would like to acknowledge the funding from MSMT LTT-20015. PK would like to acknowledge the funding from GACR 22-30516K. The funding from ANID-23-05 and HORIZON project number 101086149 enabled the mobility between Czech and Chilean colleagues (MV, PK). DI acknowledges support from the European Research Council (ERC) via the ERC Synergy Grant ECOGAL (grant 855130) and from collaborations and/or information exchange within NASA’s Nexus for Exoplanet System Science (NExSS) research coordination network sponsored by NASA’s Science Mission Directorate under Agreement No. 80NSSC21K0593 for the program ``Alien Earths”. 
\end{acknowledgements}

\bibliographystyle{aa}
\bibliography{aa}

\appendix

\section{Future atmospheric characterization: results for individual planets}
\label{atmopl}

\subsection{WASP-77 Ab}

WASP-77 Ab is listed as a Tier-3 object in a recently proposed Ariel candidate sample~\citep{ed22}, with emission spectroscopy as the preferred observing mode. \citet{line21} found a solar C/O ratio and sub-solar metallicity in the planetary atmosphere using the Gemini telescope. The sub-solar metallicity was later confirmed using the NIRSpec instrument on JWST \citep{aug23}. \citet{line21} suggested that WASP-77 Ab accreted its envelope interior to its parent protoplanetary disk’s $\rm{H_2O}$ ice line from carbon-depleted gas with little subsequent planetesimal accretion or core erosion. Such mechanisms are hard to reconcile with the non-solar abundance ratios of WASP-77 A published by \citet{reg22}. They found that WASP-77 Ab likely formed outside of its parent protoplanetary disk's $\rm{H_2O}$ ice line ($\sim$ 4.5 AU) based on enhanced C/O ratio and then migrated inwards. \citet{khor23}, using the SimAb code, studied the possible pathways for the formation of this planet with different initial disk conditions. They found that the planet has likely initiated its migration within the $\rm{CO_2}$ ice line ($\sim$ 15 AU) and was moved inward via disk-free migration. Our measurement of an apparent alignment of the planet's orbit with the stellar rotation axis is in agreement with this. Finally, \citet{chan22t} found hints of TiO in the atmosphere. Our results find a featureless spectrum that could hint either at the presence of hazes in the optical wavelengths or at a lower abundance of refractory materials. However, future studies with higher S/N are needed to confirm the origin of the muted features.

\begin{figure}[h]
\includegraphics[width=0.45\textwidth]{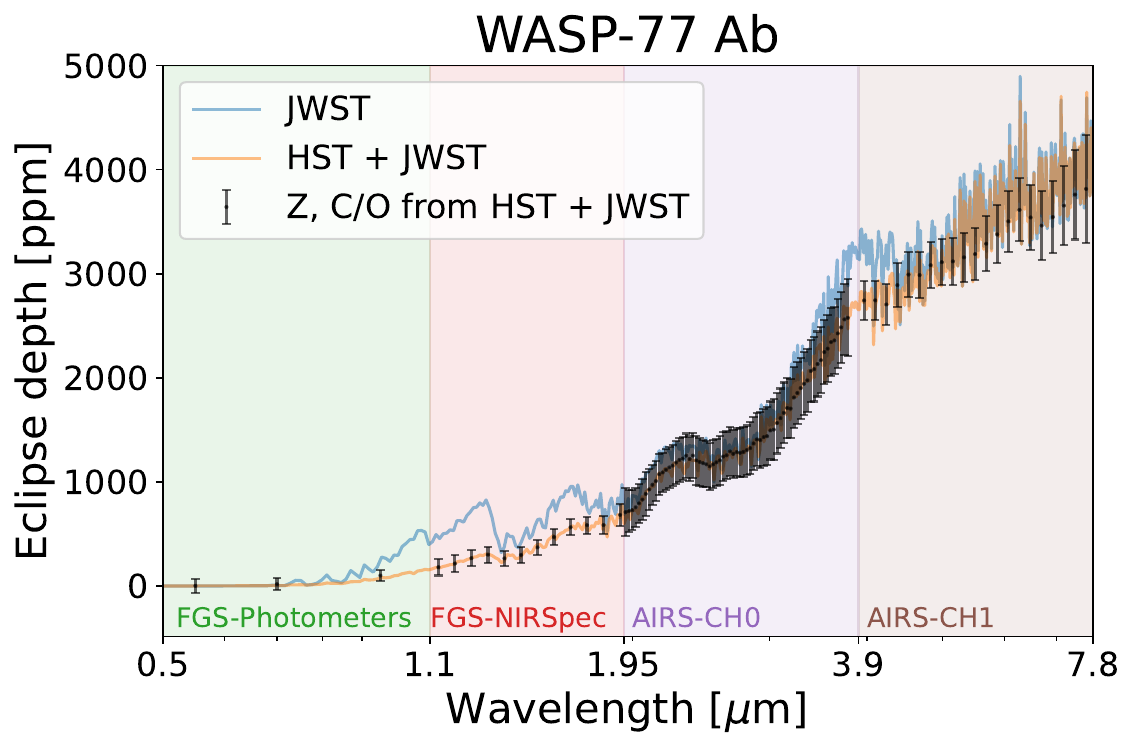}
\caption{Simulated Ariel emission spectrum of WASP-77Ab's atmosphere. In orange and blue we show two competing models presented in \citet{aug23}. Ariel will be able to distinguish between these two models with high confidence with a single observed eclipse due to its broad wavelength coverage.
}
\label{w77ts}
\end{figure}

Future characterization of WASP-77 Ab with Ariel and JWST will measure several carbon- and oxygen-bearing molecular species (such as $\rm{H_2O}$, $\rm{CO}$ and $\rm{CO_2}$) to determine its C/O ratio with unprecedented precision and accuracy to shed more light onto its formation pathways. We show a simulated spectrum using one eclipse in Fig. \ref{w77ts}. We generated the spectrum assuming equilibrium chemistry with metallicity and C/O ratio values taken from \citet{smi24}. A single Ariel observation will be able to unambiguously distinguish between the two currently competing models thanks to its continuous wavelength coverage from 0.5 to 7.8 $\mu$m observed simultaneously.

As suggested by \cite{tur21}, additional elemental ratios (e.g., N/O, C/N, and S/N) may present a better opportunity to study the planetary formation and will provide crucial information in unambiguously determining planetary history pathways. 
\citet{char22} show that WASP-77 Ab would be an excellent target for studying the redistribution of the thermal energy via obtaining phase curves during various orbital phases of the planetary orbit. Phase curves studies will cover the whole orbital phase; the night-side with temperature $\approx 1\,000$\,K \citep{keat19} offers the best opportunity to infer nitrogen abundances (in order to obtain C/N and N/O ratio). Based on detailed models by \citet{ohn22}, $\rm{NH_3}$ offers the best opportunity for detection with the JWST-MIRI instrument, targeting its strong absorption band at 11 $\mu$m.

\subsection{WASP-101b}
Due to its high TSM, WASP-101b was initially considered as a prime candidate for transmission spectroscopy using JWST in the Early Release Science (ERS) program. However, \citet{wak17101} later measured a featureless spectrum. They observed WASP-101b in the spectral region from 1.1 to 1.7$\mu$m with HST and from the lack of spectral features inferred the possible presence of a cloud deck or of water vapor depletion. This apparently disqualified WASP-101b from the JWST ERS. A later study using HST in the optical region obtained a flat spectrum and inferred the presence of clouds \citep{rat23}. In our analysis, we have obtained results consistent with the presence of clouds yielding a featureless spectrum.
Nevertheless, \citet{kaw19} have shown that, even in the presence of clouds, the redder part of the JWST or Ariel spectrum can still provide useful information about the atmospheric composition. 
WASP-101b is categorized as a Tier-3 object in~\citet{ed22}, with transmission spectroscopy as the preferred observing mode and we provide here a simulated transmission spectrum as would be observed by Ariel.

\begin{figure}[h]
\includegraphics[width=0.45\textwidth]{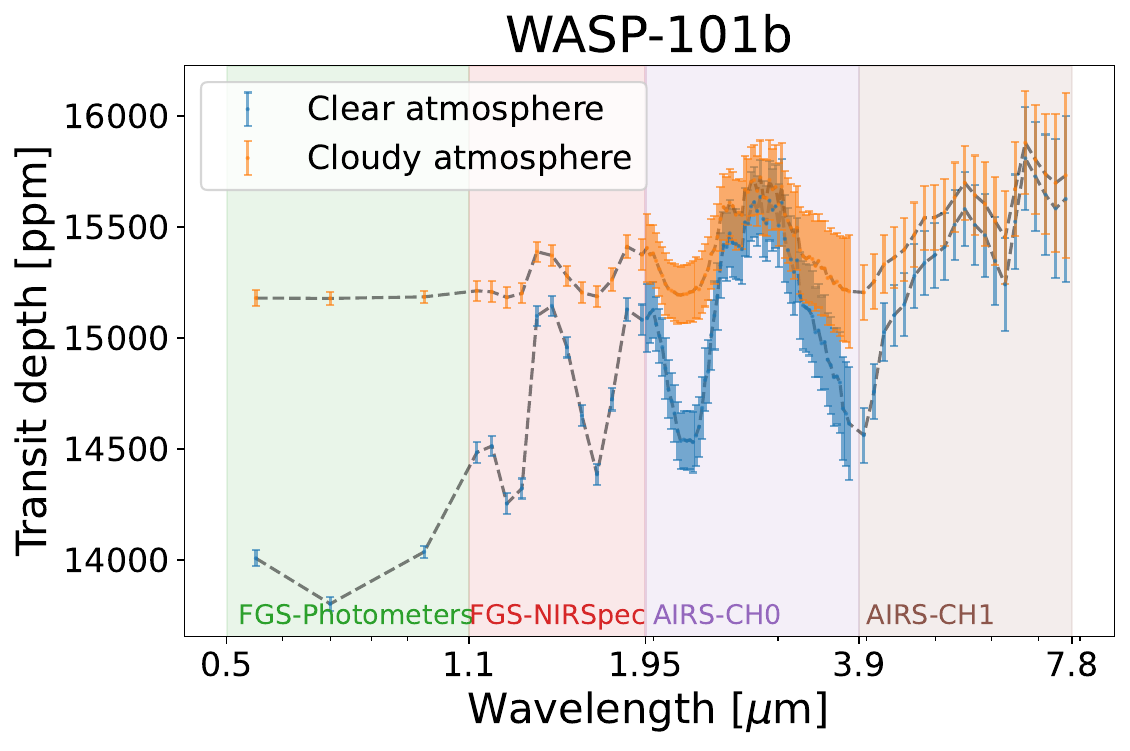}
\caption{Simulated Ariel transmission spectra of WASP-101b's atmosphere. In orange, we show a model with thick clouds consistent with previous HST data. In blue, we display a model with a clear atmosphere. Colors correspond to the different instruments/channels of the Ariel.
}
\label{w101ts}
\end{figure}

The simulated spectrum using 2 transits is presented in Fig.\,\ref{w101ts}; in blue, the spectrum corresponds to equilibrium chemistry with solar elemental ratios with a clear atmosphere. In orangee, the simulated spectrum corresponds to equilibrium chemistry and is consistent with the muted spectral features observed with HST. This shows that, even when considering gray clouds (opaque at all wavelengths), a worst-case assumption, this does not preclude the possibility of atmospheric characterization. The spectral region between 2 and 4 $\mu$m  still contains valuable information about the molecular composition -- in particular, features of CO and $\rm{CO_2}$ can still be accessible for analysis, especially as $\rm{CO_2}$ is an atmospheric metallicity indicator in giant planet atmospheres \citep{lod02}. These observations will improve our understanding of the environment where aerosols form, especially as WASP-101b might belong to a sparse class of cloudy exoplanets with equilibrium temperature above 1\,500\,K. Hence, future JWST observations with the MIRI instrument might shed more light on the possible presence, origin, and composition of the aerosols \citep{wak17c}. Furthermore, MIRI will cover the strong $\rm{SO_2}$ band between 7 and 8 $\mu$m. As $\rm{SO_2}$ is expected to be present in higher layers of the atmosphere, it is less likely to be obscured, even in the presence of aerosols. $\rm{SO_2}$ has recently been suggested as a promising tracer of metallicity in giant exoplanet atmospheres \citep{pol23,tsai23}.

\subsection{WASP-103b}

WASP-103b is ranked in the Ariel tier 3 for emission spectroscopy~\citep{ed22}. 
The atmosphere of WASP-103b has been studied by several teams in the past. \citet{len17} found the presence of strong alkali metals using the Gemini telescope. However, this result has not been confirmed by subsequent atmospheric studies \citep{wil20,kir21}. Our results are consistent with the null detection reported by the later studies. \citet{chan22} found hints of thermal inversion and presence of $\rm{H_2O}$, FeH, CO and $\rm{CH_4}$ features using HST data. Recent observations with the Canada–France–Hawaii Telescope \citep{shi23} confirmed the presence of thermal inversion and also detected super-solar metallicity composition.

\begin{figure*}[btph]
\centering
\includegraphics[width=0.75\textwidth]{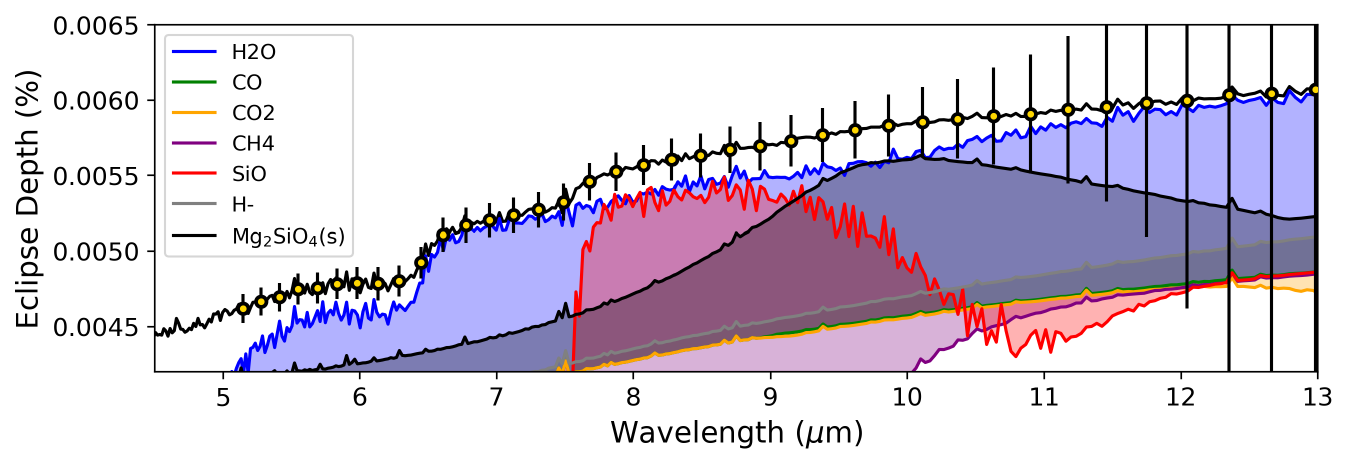}
\\ 
\includegraphics[width=0.75\textwidth]{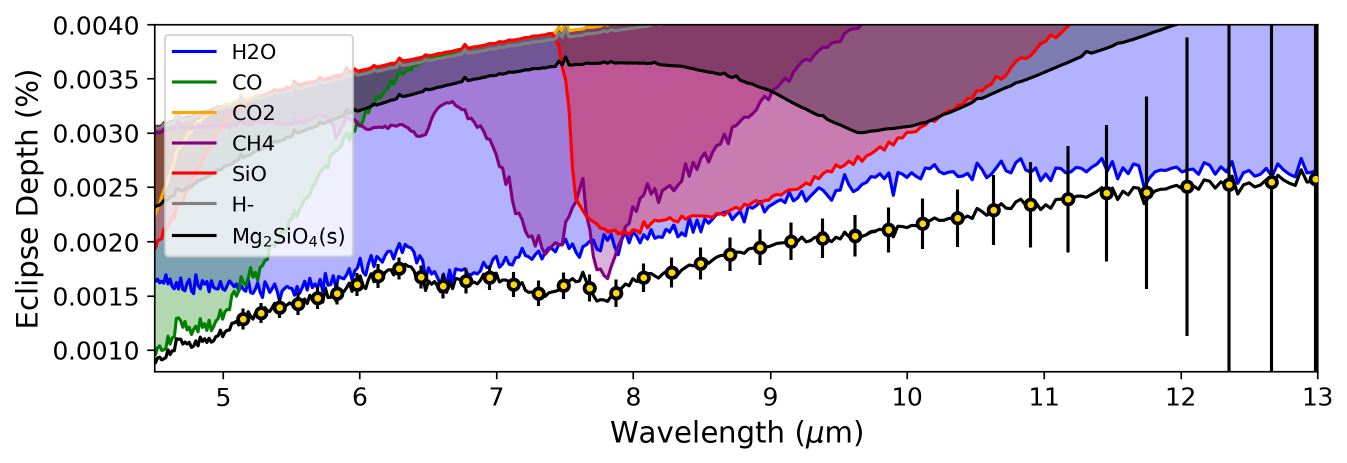}
\caption{JWST MIRI simulated spectra for day- (\textit{top}) and night-side (\textit{bottom}) of WASP-103b with phase curve observations. Yellow points mark the simulated spectra and with colored areas we show the relative contributions of various species. SiO (in red) as a tracer of refractory elements, represents a promising window into a characterization of thermal inversions in ultra-hot-Jupiters as well as their origin.}
\label{sio}
\end{figure*}

Due to its high temperature (2\,500 K) and reported detection of thermal inversion, WASP-103b is an excellent candidate for the detection of ionic species (Fe II, Ti II) and of metal oxides such as TiO and VO in the optical and SiO in the NIR \citep{gan19}. \citet{loth22} reported detection of SiO for the first time in the atmosphere of an ultra-hot Jupiter, WASP-178b, with similar equilibrium temperature ($\sim$ 2450\, K) as WASP-103b.
We used TauREx3 \citep{2021ApJ...917...37A} with the GGchem chemical code \citep{woi18} and the atmospheric parameters (T-p profiles, metallicity, and C/O ratios) inferred in \citet{chan22} to simulate day- and night-side spectrum of WASP-103b with JWST MIRI phase-curve observations (noise estimated via Pandexo: \citet{bat17}). We have included a semi-opaque silicate cloud layer in our simulations. Our models show that SiO can easily be detected for 5x solar Si/O ratio Fig. \ref{sio}. SiO is a refractory material and has been proposed as one of the possible origins of thermal inversions alongside with other refractory materials such as TiO and VO. Detection of SiO would provide key insight into the origins of thermal inversions as their origin is not fully understood \citep{parm18}. For ultra-hot-Jupiters (as they are hot enough to largely avoid the condensation of refractory species) the ratio of refractory-to-volatile elements (e.g., Si/O) combined with other ratios provides an unique insight into a planetary formation and migration as emphasized by \citet{lot21}.

Furthermore, MIRI observations may provide additional information on the condensation of silicates on the night side as the temperature should be low enough for the Si to be mostly present in the condensate form of aerosols \citep{gao20} producing a strong feature at 10 $\mu$m as found in WASP-107b \citep{dyr23}. WASP-103b hence represents an intriguing target for future JWST observations to characterize the thermal structure and origin of aerosols \citep{grant23}.  

The low density of the atmosphere with the detected tidal deformation \citep{bar22} make WASP-103 b also a good target for detection of mass loss due to Roche lobe overflow \citep{gill14} as the gravity cannot bind all the gas that escapes to the stellar environment. Roche lobe overflow has been detected in a few Jupiter-mass planets \citep[e.g.,][]{has12, yan18, cze22} using the strong hydrogen lines as such indicators are well suited for detecting mass loss \citep{wytt20}, especially in the recombination limited regime \citep{lamp23}.

Our results yield no detectable feature around the H-alpha region which is hard to reconcile with theoretical predictions \citep{gar19}. In the future, high-resolution spectrographs mounted on large telescopes (e.g., ESPRESSO and ANDES) offer the best opportunity to detect these features in the optical while space instruments such as NIRSpec/grism are best suited to detect Paschen lines in the NIR \citep{san22}. Furthermore, the NIR Helium triplet feature has been also used as a tracer of planetary mass loss \citep{sp18}, but this feature is mostly amenable for detection in systems orbiting K and late G type stars \citep{oklo19}. Recent detections of a Helium triplet in the upper atmosphere of an ultra-hot Jupiter WASP-76b \citep{casa21} and a hot Jupiter HAT-P-32 b \citep{cze22}, both orbiting F-type stars, might suggest that this feature is also accessible for detection around planets orbiting hotter stars. Helium triplet detection is now commonly done with ground-based high-resolution spectrographs. Future observations can also be done with JWST NIRISS/SOSS that has recently been shown to deliver firm He detections \citep{fu22}. The atmospheric conditions needed for atmospheric mass loss and the specific properties of the upper atmosphere are still poorly understood \citep{wo21,lamp23,bou23}.

\subsection{WASP-131b}
WASP-131b, with its high TSM, is a suitable target for further atmospheric follow-up with the transmission spectroscopy method \citep{kab19}. We detect no feature in the transmission spectrum, possibly hinting at high-altitude clouds or processes depleting sodium, but future higher S/N data are needed to confirm this.

\begin{figure}[htbp]
\includegraphics[width=0.45\textwidth]{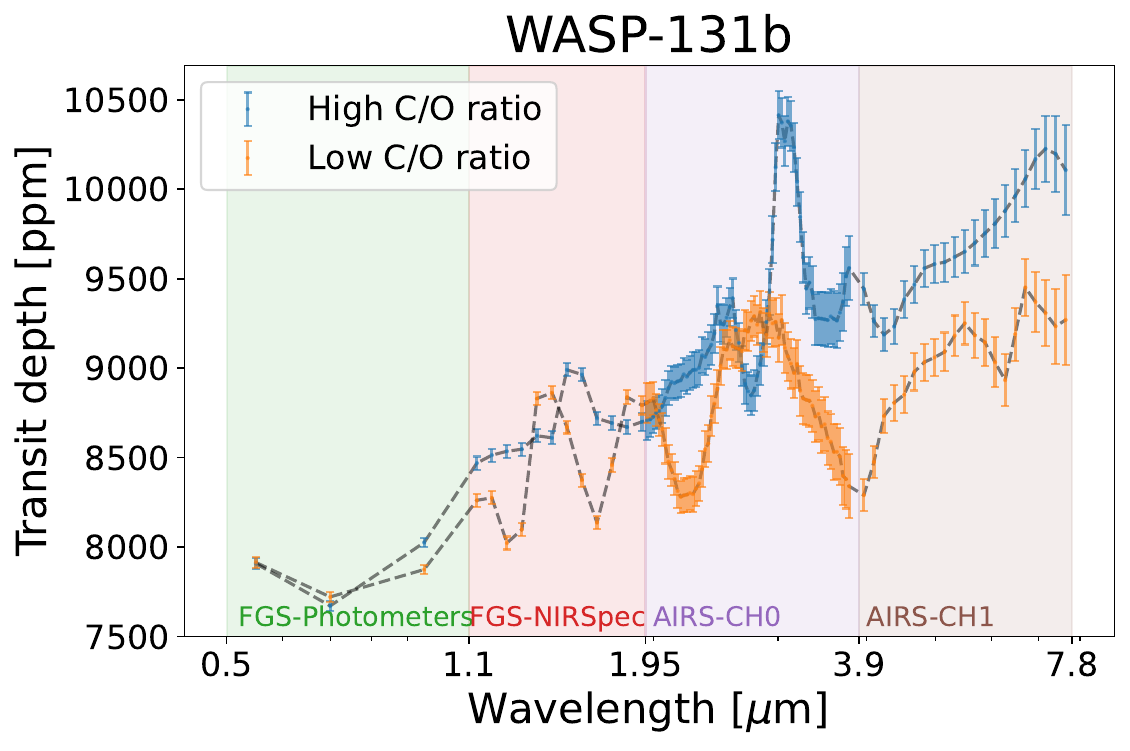}
\caption{Simulated Ariel transmission spectra of WASP-131b's atmosphere. With colors, we distinguish two models with distinct C/O ratios, corresponding to a different history of the planet.}
\label{w131ts}
\end{figure}

Fig.\,\ref{w131ts} shows the Ariel simulated spectra with the Tier 3 category, following~\citet{ed22}, using 4 transits.
The prime species for detection are $\rm{H_2O}$, CO and $\rm{CO_2}$ molecules. Such molecules will be crucial for constraining the C/O ratio, which will be able to trace the formation and migration history \citep{bre17}, especially as WASP-131b is on a highly misaligned orbit. We show in blue a spectrum with a high C/O ratio, which would hint at a formation past the ice line of the system, while in orange we show a spectrum with a low C/O ratio, hinting at a formation interior of the ice line. These two scenarios would have distinct histories that can be unraveled by atmospheric studies especially as the orbit has been already circularized by tidal forces.
~Furthermore, as suggested by \citet{tur21} using additional abundance ratios (e.g., N/O, C/N, and S/N) where possible, significantly improves the gained insights into the formation pathways of giant planets. To this account, future measurements with JWST-NIRSpec targeting $\rm{H_2S}$, may further strengthen the evolution investigation. Temperatures around 1\,500 K are quite favorable for the production of $\rm{H_2S}$. The wavelength region around 3.78 $\mu$m has been shown to be the most favorable for $\rm{H_2S}$ detection \citep{pol23}.


\newpage
\section{Additional figures}
\label{mcmcsec}
\setcounter{figure}{0}
\renewcommand{\thefigure}{B\arabic{figure}}

\begin{figure*}[h]
\includegraphics[width=\textwidth]{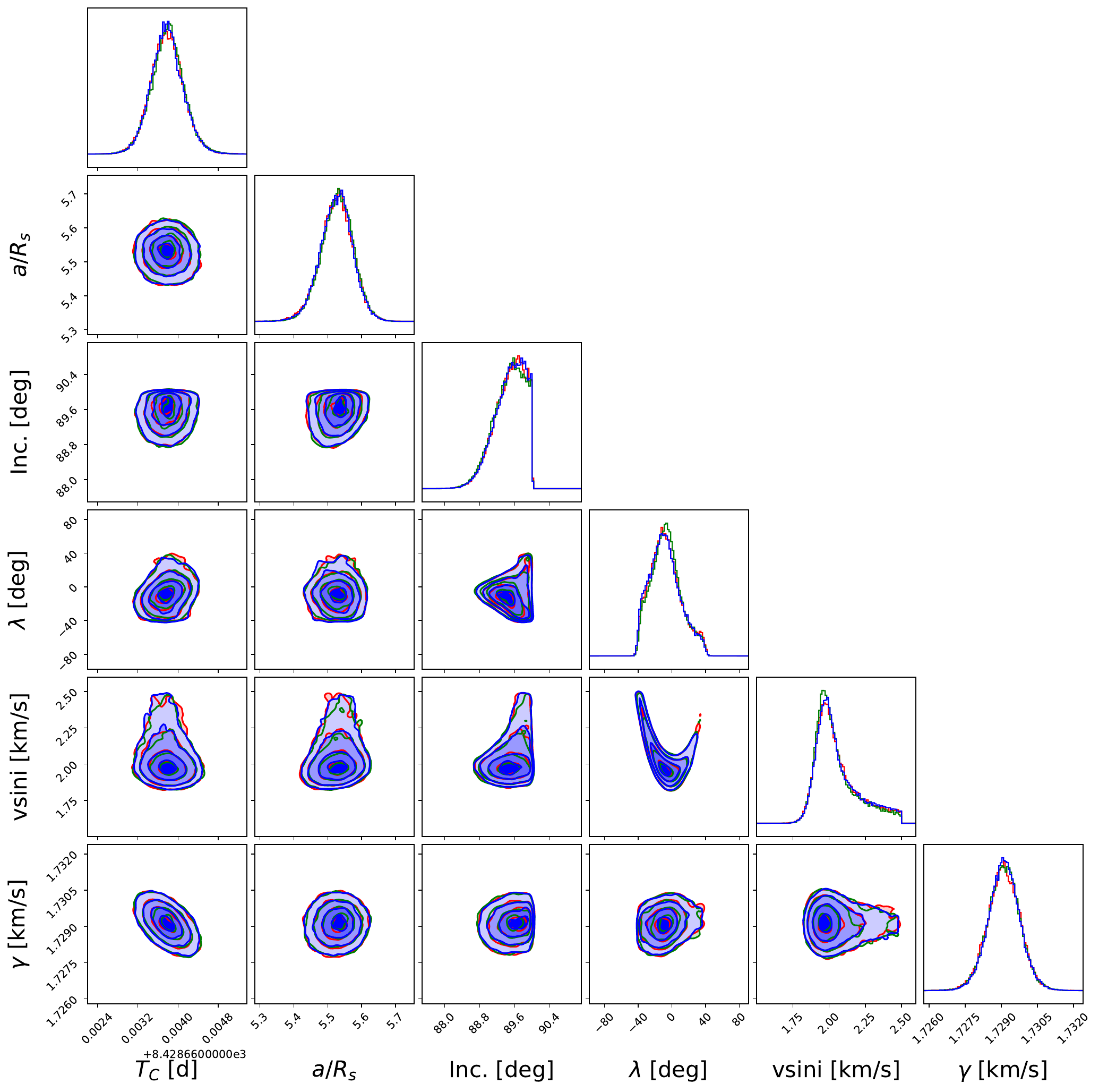}
\caption{MCMC results of WASP-77 Ab}
\label{m77}
\end{figure*}

\begin{figure*}[h]
\includegraphics[width=\textwidth]{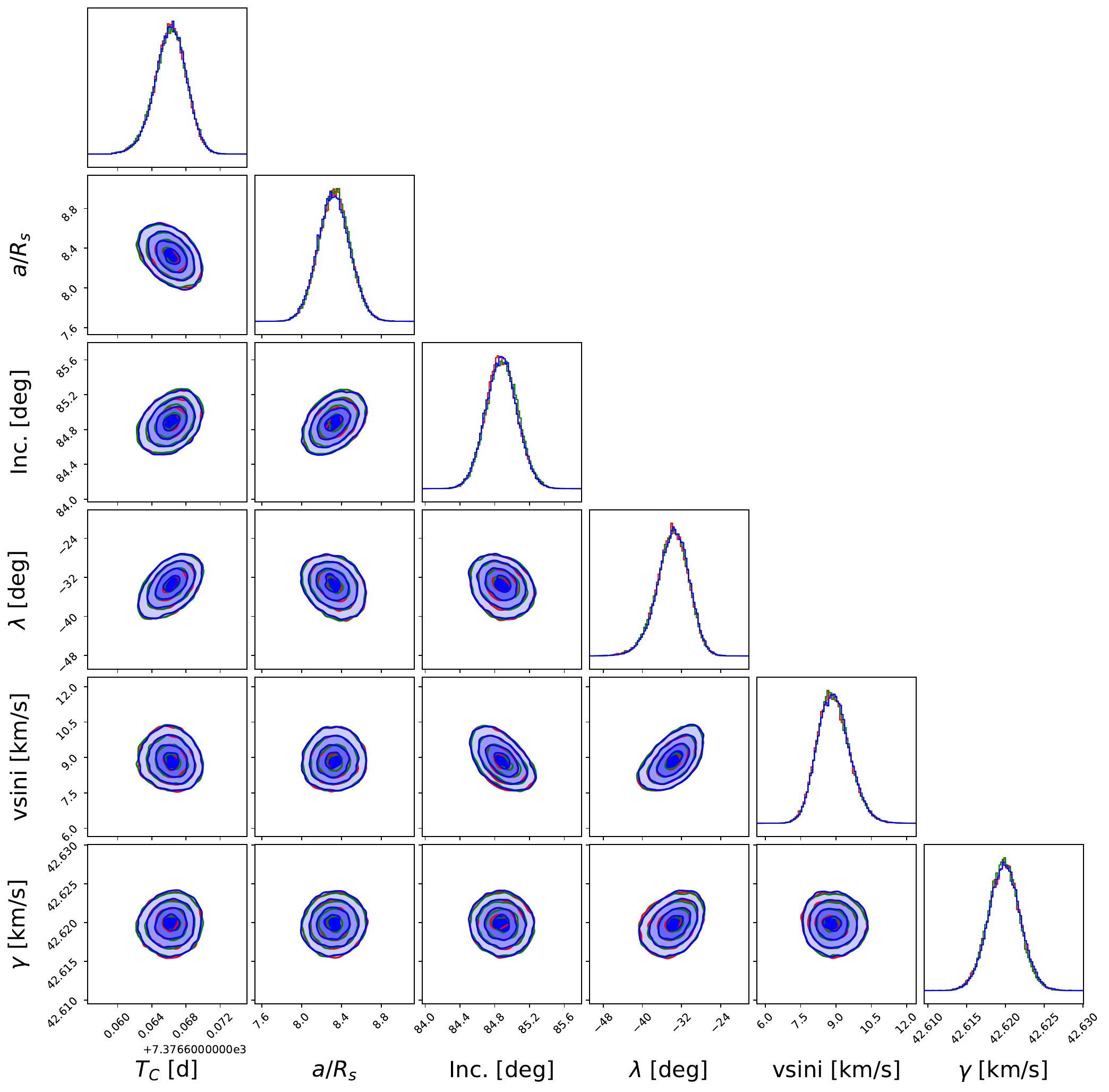}
\caption{MCMC results of WASP-101b}
\label{m101}
\end{figure*}

\begin{figure*}[h]
\includegraphics[width=\textwidth]{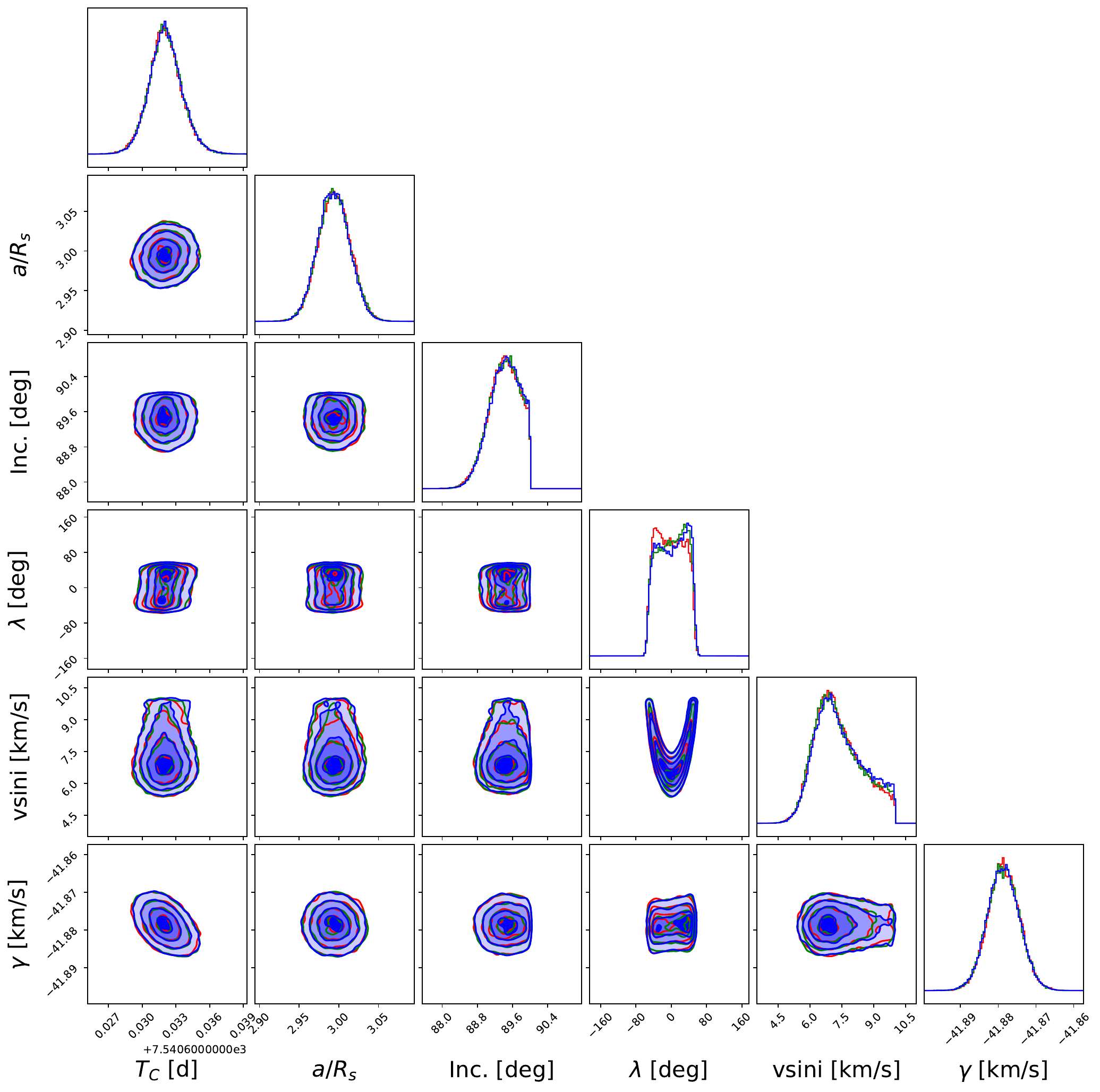}
\caption{MCMC results of WASP-103b}
\label{m103}
\end{figure*}

\begin{figure*}[h]
\includegraphics[width=\textwidth]{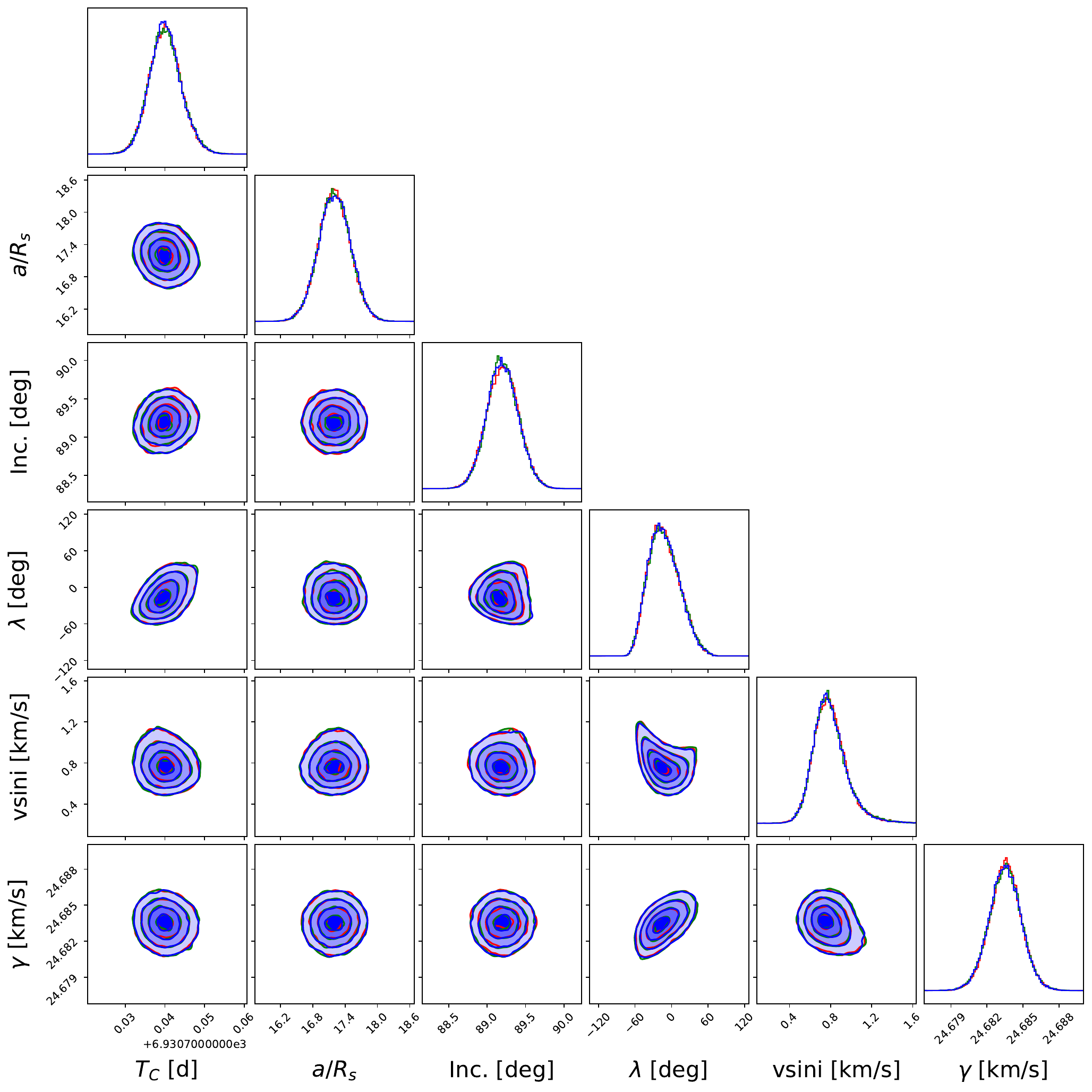}
\caption{MCMC results of WASP-105b}
\label{m105}
\end{figure*}

\begin{figure*}[h]
\includegraphics[width=\textwidth]{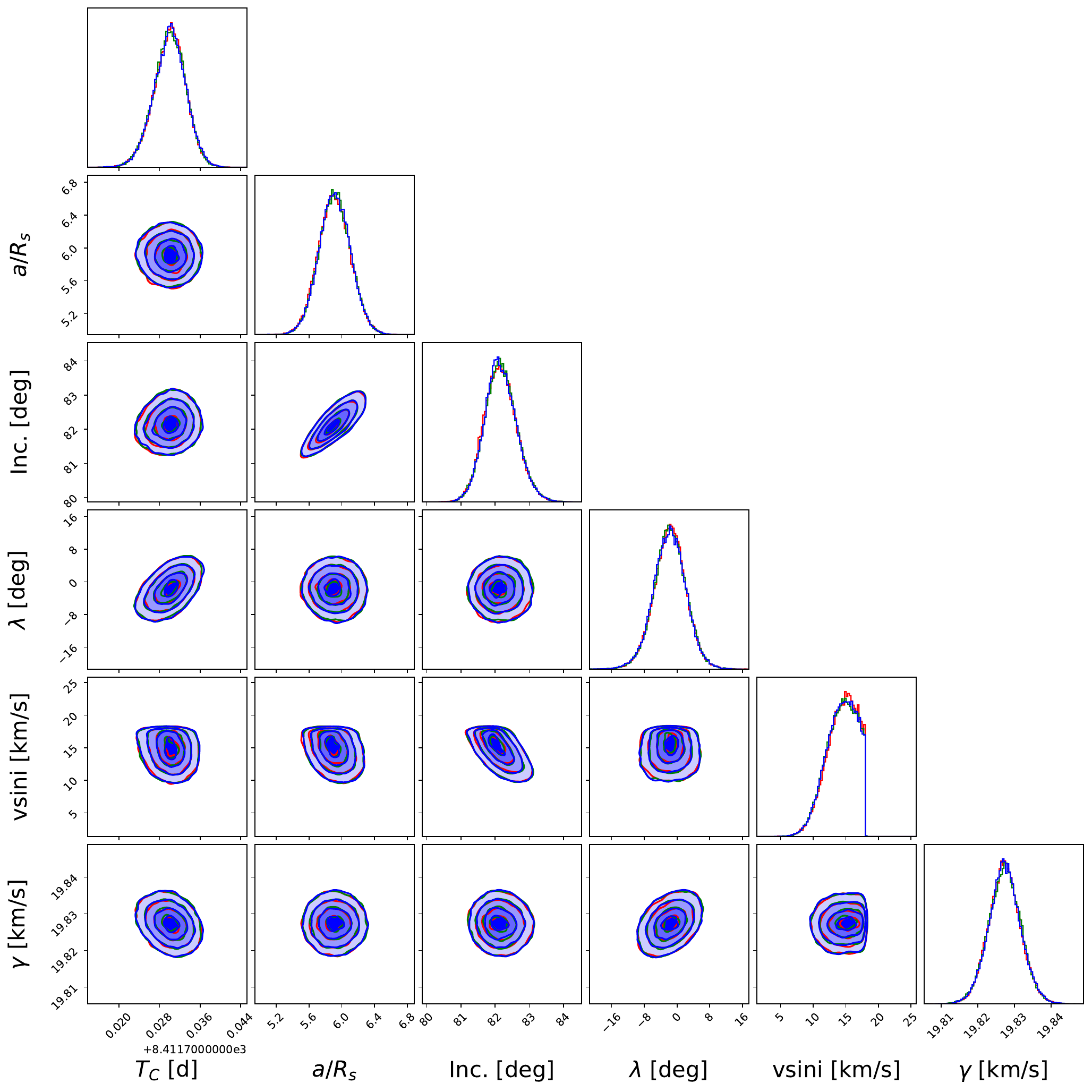}
\caption{MCMC results of WASP-120b}
\label{m120}
\end{figure*}

\begin{figure*}[h]
\includegraphics[width=\textwidth]{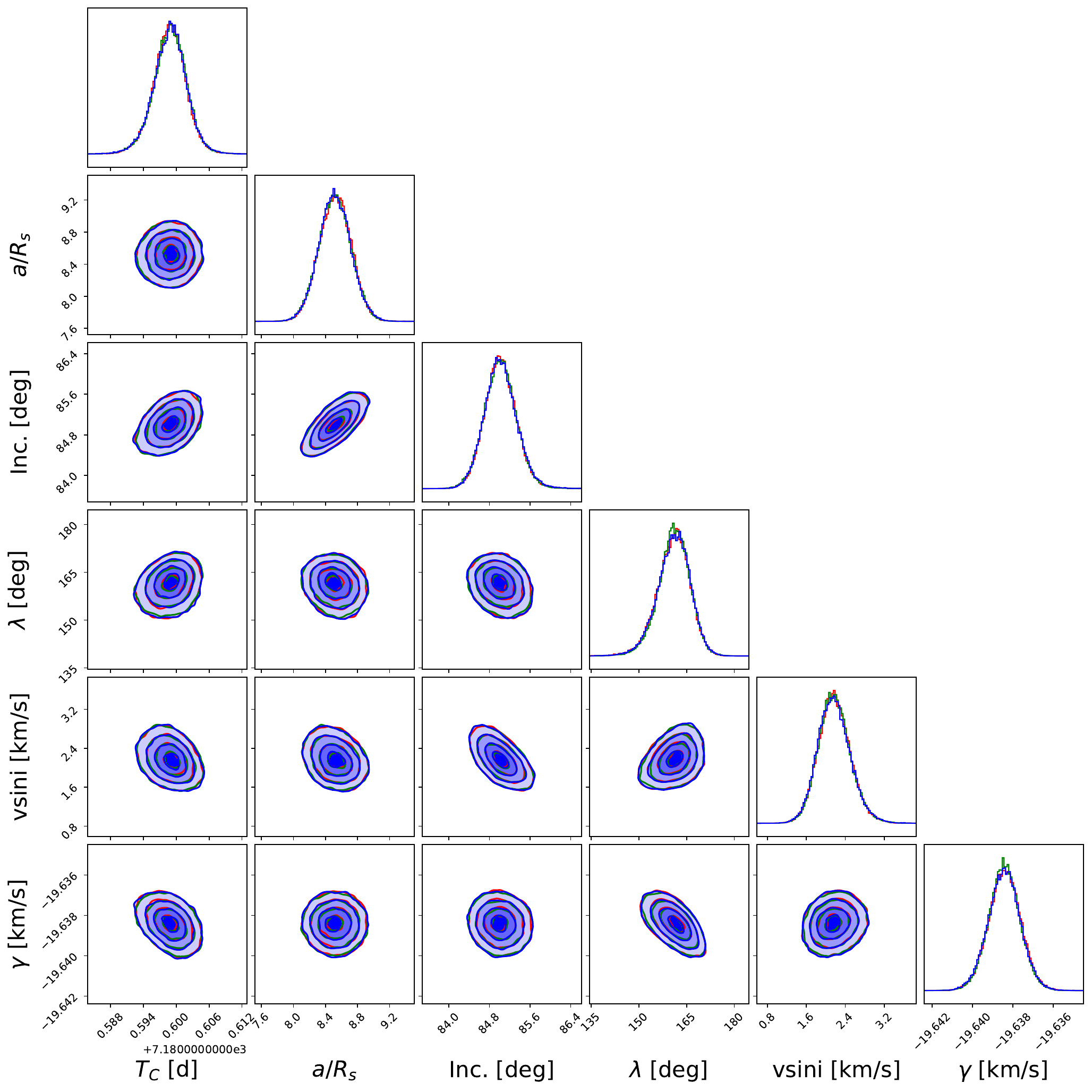}
\caption{MCMC results of WASP-131b}
\label{m131}
\end{figure*}

\begin{figure}[h]
\includegraphics[width=0.45\textwidth]{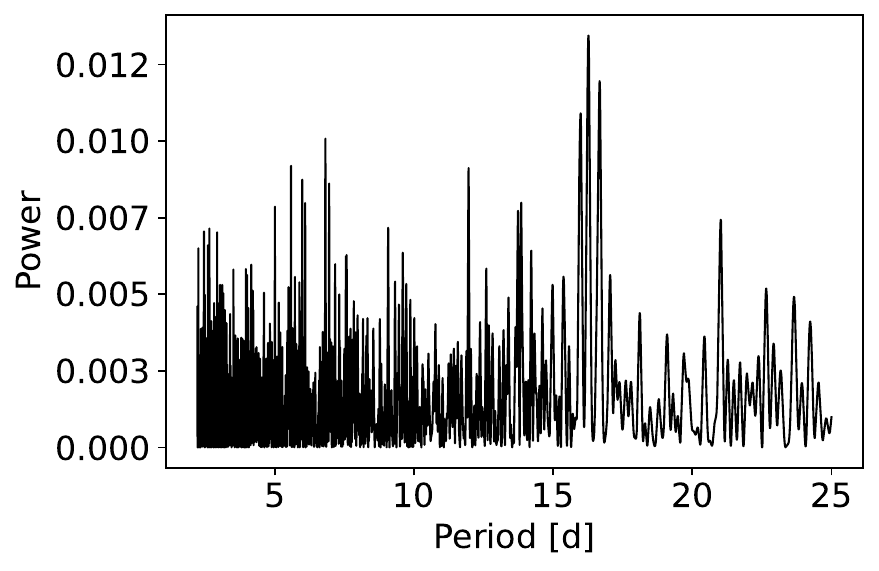}
\includegraphics[width=0.45\textwidth]{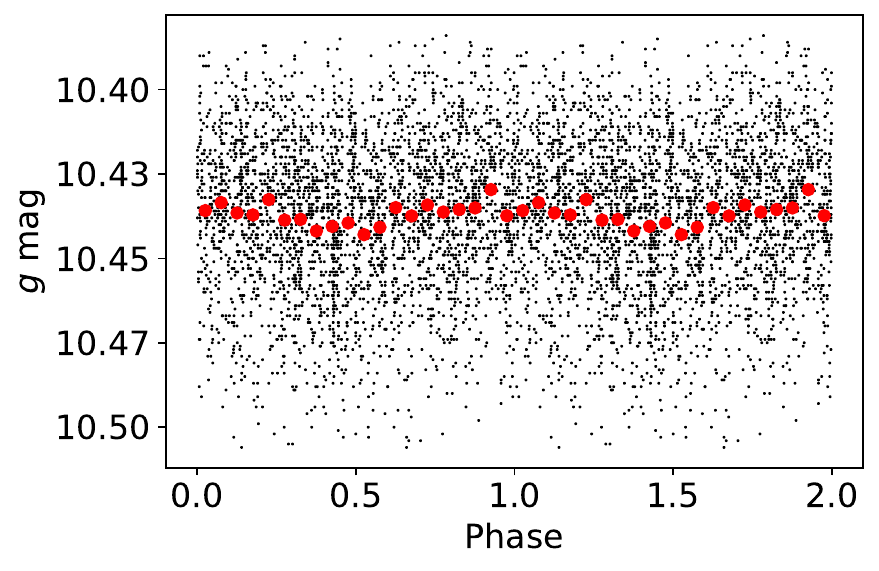}
\caption{Top: Periodogram of the ASAS-SN \textit{g} band data of WASP-77 A. We detect the strongest peak with $P_{\rm{Rot}}$ =16.2 d corresponding to the stellar rotation. Bottom: Phased ASAS-SN data with the detected stellar rotation period}
\label{fig:per}
\end{figure}


\end{document}